\documentclass[journal,transmag]{IEEEtran}

\usepackage{enumerate}

\usepackage{amssymb}

\usepackage{multirow}

\usepackage{amsmath}

\usepackage[english]{babel}

\usepackage{graphicx}

\usepackage{stfloats}

\usepackage{color}

\usepackage{amssymb}

\usepackage{bm}

\usepackage{algorithm}
\usepackage{algorithmic}

\usepackage{longtable}
\usepackage{rotating}
\usepackage{multirow}

\newtheorem{myDef}{Definition}
\newtheorem{myTheo}{Theorem}

\begin{document}
\title{Rating Protocol Design for Extortion and Cooperation in the Crowdsourcing Contest Dilemma}

\author{Jianfeng Lu,~\IEEEmembership{Member,~IEEE,}
        Yun Xin,~\IEEEmembership{}
        Zhao Zhang,~\IEEEmembership{Member,~IEEE,}
        Shaojie Tang,~\IEEEmembership{Member,~IEEE,}\\
        Songyuan Yan,~\IEEEmembership{}
        Changbing Tang,~\IEEEmembership{Member,~IEEE}


\IEEEcompsocitemizethanks{\IEEEcompsocthanksitem Jianfeng Lu is with the Department of Computer Science and Engineering, Zhejiang Normal University, Jinhua, Zhejiang, China, and also with State Key Laboratory for Novel Software Technology, Nanjing University, Nanjing, Jiangsu, China. (e-mail: lujianfeng@zjnu.cn)
\IEEEcompsocthanksitem Yun Xin, Zhao Zhang, Songyuan Yan and Changbing Tang are with the Department of Computer Science and Engineering, Zhejiang Normal University, Jinhua, Zhejiang, China. (e-mail: xinyun\_zjnu@163.com; zhaozhang@zjnu.cn; songyuanyan2560@hotmail.com; tangcb@zjnu.cn)
\IEEEcompsocthanksitem Shaojie Tang is with the Department of Information Systems, University of Texas at Dallas, Richardson, TX 75080. (e-mail:Shaojie.Tang@utdallas.edu)

}
\thanks{}}
%

\markboth{}%
{Shell \MakeLowercase{\textit{et al.}}: Bare Demo of IEEEtran.cls for IEEE Transactions on Magnetics Journals}

\IEEEtitleabstractindextext{%
\begin{abstract}
\emph{Abstract}\text{---}Crowdsourcing has emerged as a paradigm for leveraging human intelligence and activity to solve a wide range of tasks. However, strategic workers will find enticement in their self-interest to free-ride and attack in a crowdsourcing contest dilemma game. Hence, incentive mechanisms are of great importance to overcome the inefficiency of the socially undesirable equilibrium. Existing incentive mechanisms are not effective in providing incentives for cooperation in crowdsourcing competitions due to the following features: heterogeneous workers compete against each other in a crowdsourcing platform with imperfect monitoring. In this paper, we take these features into consideration, and develop a novel game-theoretic design of rating protocols, which integrates binary rating labels with differential pricing to maximize the requester's utility, by extorting selfish workers and enforcing cooperation among them. By quantifying necessary and sufficient conditions for the sustainable social norm, we formulate the problem of maximizing the revenue of the requester among all sustainable rating protocols, provide design guidelines for optimal rating protocols, and design a low-complexity algorithm to select optimal design parameters which are related to differential punishments and pricing schemes. Simulation results demonstrate how intrinsic parameters impact on design parameters, as well as the performance gain of the proposed rating protocol.
\end{abstract}

}

\maketitle

\IEEEdisplaynontitleabstractindextext

%
\IEEEpeerreviewmaketitle

\section{INTRODUCTION}
%
%
%
%

\IEEEPARstart{C}{rowdsourcing} has emerged as a new data-collection and problem-solving model, offering a distributed and cost-effective approach to obtain needed data or services by soliciting contributions from a large group of people in the online community \cite{1}. To crowdsource a task, the requester submits it to a crowdsourcing platform with an associate reward. People who can accomplish the task, called workers, can choose to work on it and devise solutions to the requester for exchanging the payment via the crowdsourcing platform. Over the past decade, techniques for securing crowdsourcing operations have been expanding steadily, so is the number of applications of crowdsourcing \cite{2}. However, workers in a crowdsourcing platform have the opportunity to exhibit antisocial behaviors due to the openness of crowdsourcing, and hence crowdsourcing is deprived of its imaginal shine when collective efforts are derailed or severely hindered by elaborate sabotage \cite{3}.

Motivated in part by the DARPA Network Challenge \cite{3a}, a crowdsourcing contest dilemma game was recently proposed in \cite{5}. It occurs a non-cooperative situation where two workers compete for a given task in a two-stage game, where the worker with a better solution wins the prize, and the loser gets nothing. In the first stage, each of these two workers determines whether he would like to devote a high level or low level effort. In the second stage, workers have the option of attacking or not attacking their opponents (\emph{e.g.}, disrupting the opponent's solution, creating multiple identities to carry out a Sybil attack, etc.) depending on whether the attack allows them to get ahead. The equilibrium analysis shows that workers will find enticement in their self-interest to free-ride by taking the payment and choosing the in-house strategy (\emph{i.e.}, low level effort) in the first stage, while in the second stage, the expected number of attacks is one, regardless of any choice of intrinsic parameters, thereby greatly reduces social utility, which is a social dilemma.

The main reason why workers in the above two-stage game have the incentive to free-ride and attack is the absence of punishments for such behaviors. On the one hand, self-interested strategic workers are inclined to adjust their strategies over time to maximize their own utilities. On the other hand, they can not receive an direct and immediate benefit by following the recommended strategy (choosing crowdsourcing and not attacking in the first and second stage, respectively). Such a conflict leads to an inevitable fact that, many workers would be apt to free-ride to take the reward but refuse to pay efforts, and attack their opponents in order to be in the lead. Therefore, the main challenge in crowdsourcing competitions is how can the requester incentivize workers to comply with the social norm, \emph{i.e.}, workers who choose the crowdsourcing strategy in the first stage and do not attack their opponent in the second stage should be rewarded immediately, otherwise, they should be punished.

Although a variety of incentive mechanisms based on pricing, reputation and reciprocity schemes have been explored to induce cooperation in crowdsourcing \cite{6,6A,6B,12B,12D,25}, existing mechanisms are not sufficiently effective due to the following features: heterogeneous workers compete against each other in a crowdsourcing platform with imperfect monitoring, and they can freely and frequently change their opponents. Hence, in order to compel selfish workers to follow the social norm and overcome the inefficiency of the socially undesirable equilibrium, it is of great importance to design optimal incentive mechanisms by taking these features into consideration.

In this paper, we aim to develop a novel game-theoretic design of rating protocols to address the crowdsourcing contest dilemma. The main topic of this paper is to maximize requesters' utilities by extorting selfish workers and enforcing cooperation among them, \emph{i.e.}, paying workers as little as possible while providing sufficient incentives for individual workers to follow the social norm in order to sustain high-performance crowdsourcing platform, and thus evade socially undesirable equilibrium. To the best of our knowledge, standing from the requester's point of view to maximize his utility in all time periods is rarely studied by other works. We believe studies on this topic is urgent since cost-efficiency is one of the main attractions of crowdsourcing, requesters may not have enough incentive to post tasks via a crowdsourcing platform if they cannot earn enough benefit. In this paper, we analyze how cooperation can be enforced and how to maximize the utility of a requester under the designed rating protocol by extorting selfish workers and enforcing cooperation among them. Our work is based on game theory because it has been found to be a powerful tool to study strategic interactions among selfish and rational individuals and design incentive mechanisms to stimulate cooperation among them \cite{25A,tdsc1,tmc1}.

\subsection{Main Contributions}
The following is a list of our main contributions.

\begin{itemize}
  \item  Standing at the protocol designer's point view, we explore the strategy of requesters aiming to maximize their utilities on all tasks and provide workers sufficient incentives of contributing good behaviors in order to sustain high-performance crowdsourcing. To the best of our knowledge, this is the first work achieving extortion and cooperation simultaneously in crowdsourcing competitions.
  \item  Workers' heterogeneity is taken into consideration when designing utility functions of a worker in the two-stage game, and thus we model the crowdsourcing contest dilemma as an asymmetric game. This makes our rating protocol applicable in heterogeneous crowdsourcing platforms.
  \item  A novel game-theoretic design of rating protocols that integrate binary rating labels with differential pricing is developed to incentivize workers to contribute good behaviors. Our rating protocol can achieve the social optimum, which is easy to design and flexible to implement.
  \item Differential punishments are used to transfer payoffs from low-rating workers to high-rating workers, which can reduce performance loss in the presence of imperfect monitoring while providing sufficient incentives for individual workers to follow the social norm.
  \item The problem of designing an optimal rating protocol that maximizes the revenue of the requester among all sustainable rating protocols is formulated, we rigorously analyze how heterogeneous workers' behaviors are influenced by intrinsic parameters and design parameters as well as the workers' evaluation of their individual long-term utilities, in order to characterize the optimal design.
  \item A low-complexity algorithm is proposed to select optimal design parameters which are related to differential punishments and pricing schemes. Simulation results show the validity and effectiveness of our proposed algorithm for crowdsourcing contest dilemma.

\end{itemize}

\subsection{Related Work}
In recent decades, it has already been noticed that there is an urgent need to stimulate cooperation among self-interested workers in crowdsourcing by introducing incentive mechanism \cite{6A,6B,tmc2}. There exist many types of incentive mechanisms such as \emph{pricing} and \emph{reputation} \cite{6C}. Incentive mechanisms based on pricing incentivize individuals to provide good behaviors relying on monetary or matching rewards in the form of micropayments, which in principle can achieve the social optimum by internalizing external effects of self-interested individuals \cite{7,8}. However, as pointed out by \cite{12B}, \cite{8A} and \cite{12}, if an inefficient pricing based incentive mechanism is applied, ``free-riding'' happens when rewards are paid before a task starts, a worker always has the incentive to take the reward but refuse to devote efforts, whereas if rewards are paid after the task is complete, ``false-reporting'' arises since the requester has the incentive to lower or refuse the reward to workers by lying about the outcome of the task. Incentive mechanisms based on reputation scheme, on the other hand, take individuals' reputation into consideration, which reward and punish individuals according to individuals' past behaviors \cite{12D,12C,13,tmc3}. A central reputation entity can offer a robust method to sustain cooperation \cite{13A}. However, such an approach needs necessary traffic monitoring mechanisms, which will put a great burden on the central entity, and make it impractical in a large community. Alternatively, a distributed adaptive reputation scheme was proposed to provide a dynamical updating of reputation \cite{14}. The distributed reputation scheme, which does not rely on a central bank to control the currency, involves more complicated reputation update \cite{15}.

Traditional pricing and reputation schemes being used separately may be inefficient in a crowdsourcing contest dilemma game in which workers are part of a community and repeatedly interact. This is because workers' behaviors are influenced by incurred costs and designed payment, as well as their long-term utilities, which cannot be solely determined by a pricing scheme. Besides, workers choose to crowdsource and devise solutions in exchange for payment, increasing workers' reputation without differential payment cannot decrease their malicious behaviors. Recently, a considerable amount of efforts have been devoted using game theory to analyze how to maximize the social welfare while enforcing cooperation among individuals under a designed incentive mechanism \cite{tmc1,16,17,18,19}. Such incentive mechanisms are based on the principle of \emph{reciprocity} and can be classified into \emph{direct reciprocity} and \emph{indirect reciprocity} \cite{20}. In a direct reciprocity mechanism, individuals can identify each other, and behaviors between them are based on their personal experience with each other. Direct reciprocity mechanism is highly effective in sustaining cooperation in a small system where individuals can identify each other and interact frequently with fixed partners \cite{21}. However, in most crowdsourcing platforms, workers have asymmetric service requirements and they can freely and frequently change their partners. Hence a personal history of past reciprocation with the same partner cannot be established. To encourage self-interested workers to provide good behaviors in crowdsourcing platforms, indirect reciprocity solutions have been proposed \cite{17,22}, in which individuals decide their actions based on the available information including indirect information. Hence, an individual can be rewarded or punished by other individuals in a crowdsourcing platform even they have not had past interactions with him \cite{23}.

To implement indirect reciprocity in crowdsourcing, it is important to share as little as possible but enough amount of information about past interactions in a platform. The use of rating labels as a summary record of an individual requires significantly less amount of information being maintained, hence, the rating based incentive mechanism has a potential to form a basis for successful incentive mechanisms in a crowdsourcing platform. M. Kandori original proposes a rating protocol for a large anonymous society \cite{24}, in which each individual is attached a rating label based on the individual's past behaviors indicating his social status in the system. Under a rating protocol, the rating label of an individual who complies with (\emph{resp.} deviates from) the social norm goes up (\emph{resp.} down), and individuals with different rating labels are treated differently by the other individuals they interact with. Hence, an individual with high/low rating label can be rewarded/punished by other individuals in the crowdsourcing platform who have not had past interactions with him. Recently, a variety of rating protocols have been explored to force cooperation in crowdsourcing platforms \cite{25,23,26,27,28}. However, as shown in \cite{6}, several factors hinder the direct implementation of these works in crowdsourcing competitions. These factors can be summarized as: (\emph{i}) competitive relations exist not only between workers, but also between workers and requesters; (\emph{ii}) In the presence of imperfect monitoring, individual's rating may be inaccurate; (\emph{iii}) workers can freely and frequently change their opponents in the two-stage game. Because of these features of crowdsourcing competitions, it is important to overcome the inefficiency of the socially undesirable equilibrium by designing a novel rating protocol.

In our previous work \cite{6}, we have designed a rating protocol based on game theory to address the crowdsourcing contest dilemma. In such a rating protocol, we capture the fundamental aspect of providing necessary and sufficient incentives for workers to contribute good behaviors. Yet, it only considered the expected one-period utility. As a result, incentives for requesters to crowdsource task will be greatly reduced or eliminated, because no requester will choose to crowdsource tasks if they pay more than they earn. Therefore, from the protocol designer's point of view, it is crucial to maximize requesters' utilities while sufficient incentives are provided for workers to contribute good behaviors in order to sustain high-performance in the crowdsourcing. Besides, our previous work assumes that workers are homogenous, which is a strong assumption as many crowdsourcing scenarios are heterogeneous in the sense that workers' abilities are different. Therefore, in order to apply the designed rating protocol to heterogeneous crowdsourcing platforms, it is necessary to take workers' heterogeneity into consideration when designing utility functions.

\subsection{Paper Organization}
The remainder of this article is organized as follows. In section II, we describe the crowdsourcing contest dilemma game with rating protocols. In section III, we formulate the problem of designing an optimal rating protocol with constraints. Then we design the optimal rating protocols in Section IV.  Section V presents simulation results to illustrate key features of the designed rating protocol. Finally, conclusions are drawn in Section VI.

\section{SYSTEM MODELS}

\subsection{System Setting}
\begin{table}[tb]
\scriptsize
\caption{SUMMARY OF NOTATIONS IN THIS PAPER}
\label{table1}
\begin{center}
\begin{tabular}{|c|c|c|c|c|c|c|}
\hline

  \multicolumn{1}{c|}{\textbf{Notations}}& \multicolumn{6}{|c}{\textbf{Physical Meanings}} \\
\hline
\multicolumn{1}{c|}{$c_i$}& \multicolumn{6}{|l}{cost of the \emph{i}th-worker in the first stage.}\\
\hline
\multicolumn{1}{c|}{$s_i$}& \multicolumn{6}{|l}{cost of the \emph{i}th-worker in the second stage.}\\
\hline
\multicolumn{1}{c|}{$d$}& \multicolumn{6}{|l}{damage inflicted by an attack.}\\
\hline
\multicolumn{1}{c|}{$\varepsilon_i$}& \multicolumn{6}{|l}{probability that errors occur in the \emph{i}th-stage game.}\\
\hline
\multicolumn{1}{c|}{$\delta$}& \multicolumn{6}{|l}{discount factor to denote workers' patience.}\\
\hline
\multicolumn{1}{c|}{$\alpha$}& \multicolumn{6}{|l}{The strength of reward imposed to workers.}\\
\hline
\multicolumn{1}{c|}{$\beta$}& \multicolumn{6}{|l}{strength of punishment imposed to workers.}\\
\hline
\multicolumn{1}{c|}{$\gamma_\theta$}& \multicolumn{6}{|l}{payment rewarded to $\theta$-worker.}\\
\hline
\multicolumn{1}{c|}{$\mathcal P$}& \multicolumn{6}{|l}{rating protocol.}\\
\hline
\multicolumn{1}{c|}{$\theta$}& \multicolumn{6}{|l}{rating label.}\\
\hline
\multicolumn{1}{c|}{$\Theta$}& \multicolumn{6}{|l}{set of rating labels.}\\
\hline
\multicolumn{1}{c|}{$\sigma$}& \multicolumn{6}{|l}{social strategy.}\\
\hline
\multicolumn{1}{c|}{$\tau$}& \multicolumn{6}{|l}{rating scheme.}\\
\hline
\multicolumn{1}{c|}{$\psi$}& \multicolumn{6}{|l}{pricing scheme.}\\
\hline
\multicolumn{1}{c|}{$v$}& \multicolumn{6}{|l}{expected one-period utility of a worker.}\\
\hline
\multicolumn{1}{c|}{$v^\infty$}& \multicolumn{6}{|l}{expected long-term utility of a worker.}\\
\hline
\multicolumn{1}{c|}{$u$}& \multicolumn{6}{|l}{expected one-period utility of a requester.}\\
\hline
\multicolumn{1}{c|}{$U_{\mathcal P}$}& \multicolumn{6}{|l}{social utility under the rating protocol $\mathcal P$}\\
\hline
\end{tabular}%
\end{center}
\end{table}

In a crowdsourcing contest dilemma game \cite{5,6}, two competing workers interact with each other in a two-stage game to obtain a better solution to a task which can be crowdsourced. In the first stage, each of these two workers can choose to achieve a given task via crowdsourcing (denoted as \emph{C}) or solve the problem in-house (denoted as \emph{S}). As any \emph{C} strategy is costly, and workers' costs are different by considering their heterogeneity, we assume that the \emph{i}th-worker will consume a cost $c_i\in(0,1)$ for selecting \emph{C}, while the cost is approximated to 0 if choosing \emph{S}. In the second stage, workers decide whether or not to attack their opponents (attacking is denoted by \emph{A}, while not attacking is denoted by \emph{N}). Similarly, an attack is costly, we assume that the \emph{i}th-worker will consume cost $s_i\in(0,1)$ to attack his opponent, in order to inflict the damage $d\in(0,1)$ (the attacking process is socially valuable only if $s_i \le d$). Each one of $c_i$, $s_i$ and $d$ can be referred to as a fraction of the total reward $\gamma$, which is normalized to be 1. The discount factor $\delta$ is the rate at which a worker discounts his future payoff and reflects the patience of him. Taking into account imperfect monitoring, the outcome of the task received by the requester is inconsistent with the effort of the worker. Let $\varepsilon_1$ and $\varepsilon_2$ denote the probabilities that monitoring or reporting errors occur in the first stage and the second stage, respectively. In short, there exist eight intrinsic parameters in a crowdsourcing contest dilemma game, namely $c_1$, $c_2$, $s_1$, $s_2$, $d$, $\delta$, $\varepsilon_1$ and $\varepsilon_2$. We summarize these parameters as well as other notations used in the paper in Table \ref{table1}.

In the above proposed model, workers' heterogeneity (\emph{i.e.}, workers¡¯ costs and abilities are different) is taken into consideration, which is the only difference from our previous one \cite{6}. This makes our model more close to reality. The entire pay-off matrix for the revised game played in the first stage was depicted in Table \ref{table2}, where we set $\mathcal X_i=1/2-s_id+(s_id^2)/2$, $\mathcal Y_i=[d-(c_i+d)^2/2](1-c_i-s_i)+(1-c_i-d)^2(1-c_i)/2$). The concrete computation process for Table \ref{table2} is given in Appendix A.

\begin{table}[tb]
\caption{THE EXPECTED PAYOFF MATRIX FOR THE FIRST-STAGE}
\label{table2}
\begin{center}
\begin{tabular}{cc|c|c|c|c|c|c|}

  \multicolumn{2}{c}{\textbf{$$}}& \multicolumn{6}{c}{$worker$ 2} \\
  \multicolumn{2}{c}{\textbf{$$}}& \multicolumn{3}{c}{\textbf{$C$}}& \multicolumn{3}{c}{\textbf{$S$}} \\
\cline{3-8}
\multirow{2}*{$worker$ 1} & \multicolumn{1}{c}{\textbf{$C$}}& \multicolumn{3}{|c|}{$\mathcal X_1-c_1/2,\mathcal X_2-c_2/2$}& \multicolumn{3}{c|}{$\mathcal Y_1,0$} \\
\cline{3-8}
\multicolumn{1}{c}{\textbf{$$}}& \multicolumn{1}{c}{\textbf{$S$}}& \multicolumn{3}{|c|}{$0,\mathcal Y_2$}& \multicolumn{3}{c|}{$\mathcal X_1,\mathcal X_2$} \\
\cline{3-8}
\end{tabular}%
\end{center}
\end{table}

In the resulting equilibrium, for any choice of intrinsic parameters, only \emph{CC} and \emph{SS} can be pure strategy equilibria in the first stage, and malicious behavior is the normal, not the abnormal, \emph{i.e.}, the expected number of attacks is one regardless of any choice of intrinsic parameters, which is contrary to the conventional wisdom in the area. In other words, workers find enticement to free-ride and attack in their self-interest, an inefficient outcome arises for such a myopic equilibrium.

\subsection{Rating Protocol}
In order to overcome the inefficiency of the socially undesirable equilibrium by extorting selfish workers and enforcing cooperation among them, \emph{i.e.}, taking the requesters' point of view and trying to maximize their average utilities on all tasks, while provide sufficient incentives to compel rational and selfish individuals to contribute good behaviors in tasks, we devote with rating protocol to incentivize self-interested workers to comply with the social norm, and thus evade the myopic equilibrium. In this paper, we integrate binary rating labels with differential pricing to incentivize workers to contribute good behaviors. In order to provide enough incentives by transferring payoffs from low-rating workers to high-rating workers, we use differential punishments that punish workers with different ratings differently. The proposed rating protocol is defined as follows.

\begin{myDef}
A rating protocol $\mathcal P$ is defined as the rules that a crowdsourcing platform uses to regulate the behavior of his workers, and is represented as a quadruple $(\Theta,\sigma,\tau,\psi)$, \emph{i.e.,} a set of binary rating labels $\Theta$, a social strategy $\sigma$, a rating scheme $\tau$, and a pricing scheme $\psi$.
\begin{itemize}
       \item 	$\Theta=\{0,1\}$ denotes the set of binary rating labels, where 1 is the good rating and 0 is the bad rating.

      \item    $\sigma:\Theta \rightarrow \mathcal{A}$ represents the adopted social strategy for a worker with rating label $\theta$, where $ \sigma(\theta|\theta \in \Theta )\in \{CN,CA,SN,SA\}$.

      \item  	$\tau :\Theta \times \mathcal{A}\rightarrow \Delta(\Theta)$ specifies how a worker's rating should be updated based on his adopted strategies and current rating as follows:
\begin{eqnarray}{\tau (\theta'| \theta ,\sigma (\theta) )=}
      \left \{
\begin{array}{ll}
  1, &\emph{if} ~\theta'=1, \theta=1 \\&\emph{and}~  \sigma (\theta)=\emph{CN} \\
  \alpha, & \emph{if} ~\theta'=1, \theta=0 \\&\emph{and}~  \sigma (\theta)=\emph{CN} \\
  1-\alpha, & \emph{if} ~\theta'=0, \theta=0 \\&\emph{and}~  \sigma (\theta)=\emph{CN} \\
  1, & \emph{if} ~\theta'=0, \theta=0 \\&\emph{and}~  \sigma (\theta)\neq \emph{CN} \\
  \beta, & \emph{if} ~\theta'=0, \theta=1 \\&\emph{and}~  \sigma (\theta)\neq \emph{CN} \\
  1-\beta, & \emph{if} ~\theta'=1, \theta=1 \\&\emph{and}~  \sigma (\theta)\neq \emph{CN}
\end{array}
\right.
\end{eqnarray}

          \item 	$\psi:\Theta \rightarrow R $ defines the rules that rewarding/punishing workers by implementing differential prices for the contributions according to the rating of workers:
          \begin{equation} {}
\psi(\theta)=\gamma_\theta,~\forall \theta \in \Theta
\end{equation}
where $\gamma_0$ is the minimal price, $\gamma_1$ is the maximal price.
 \end{itemize}
\end{myDef}
\begin{figure}
\centering
\includegraphics[height=3.42cm,width=7.2cm,angle=0,scale=1]{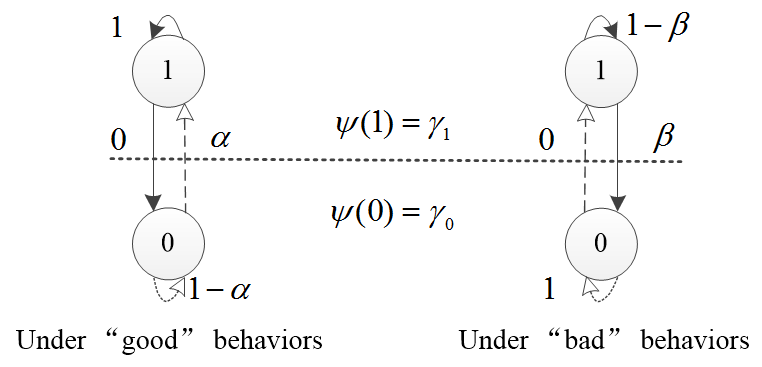}
\caption{Schematic representation of the proposed rating protocol $\mathcal P$}
\end{figure}

\emph{\textbf{Remark}}: A schematic representation of a rating scheme $\tau$ based on Definition 1 is provided in Figure 1. Under the rating update rule, if the social strategy adopted by a worker with rating 0 is observed to be \emph{CN} (\emph{i.e.}, the worker chooses \emph{C} in the first stage game and chooses \emph{N} in the second stage game.), his rating will increase to 1 with probability $\alpha$, and hold rating 0 with probability $1-\alpha$; otherwise, if the social strategy adopted by a worker with rating 0 is not \emph{CN}, he will hold rating 0 with probability 1. The analysis for the scenario of a worker with rating 1 follows in a similar manner. Hence, $\alpha$ can be referred to as the strength of reward imposed on workers when they contribute good behaviors, while $\beta$ can be referred to as the strength of punishment imposed on workers when they do not follow the recommended strategies. Other more elaborated rating update rules may be considered, but we will show that this simple one is good enough to enforce cooperation among self-interested workers, and maximizes the social utility of the requester. In order to enforce incentives on service provision, the protocol designer assigns workers with high rating higher payment as reward and hence, workers are encouraged to contribute good behaviors to increase their ratings for a return of higher payment. To determine the range of feasible prices $\gamma_0$ and $\gamma_1$, we assume that the maximum benefit which a requester obtains from receiving one unit of service is 1. It is obvious that the feasible prices should be restricted to $0\le\gamma_0<\gamma_1\le 1$.

\section{PROBLEM FORMULATION}

\subsection{Stationary Rating Distribution}
Given a rating protocol $\mathcal P$, suppose that each worker always follows the given recommended strategy $\sigma_{\mathcal P}(\theta)=CN,\forall \theta \in \Theta$, who is called as a ``compliant worker'', where the worker who deviates from social norm and plays $\sigma'(\theta)\in\{CA,SA,SN\}$ is called as a ``non-compliant worker''. As time passes, the ratings of compliant workers are updated, hence the distribution of ratings in a crowdsourcing platform evolves as a Markov chain, whose transition probability $p_{\mathcal P}(\theta'|\theta)$ is determined by the recommended strategy employed by workers. Since we are interested in the long term utilities of workers, we study the rating distribution in the long run when all workers follow the given recommended strategy $CN$. Let $\{\eta_{\mathcal P}^t(\theta)|\theta\in\Theta\}$ be the fraction of $\theta$-workers (\emph{i.e.}, a worker with rating $\theta$ is denoted by $\theta$-worker) in the total population at the beginning of an arbitrary period \emph{t}, then the transition from $\{\eta_{\mathcal P}^t(\theta)\}_{\theta=0}^{1}$ to $\{\eta_{\mathcal P}^{t+1}(\theta)\}_{\theta=0}^{1}$ is determined by the rating scheme $\tau$, as shown in Eq. (\ref{eq16}).
\begin{equation}{}
\label{eq16}
\left\{
\begin{array}{ll}
\eta_{\mathcal P}^{t+1}(0)=&p_{\mathcal P}(0|1,\sigma (1)=CN)\eta_{\mathcal P}^{t}(1)\\&+p_P(0|0,\sigma (0)=CN)\eta_{\mathcal P}^{t}(0)\\
\eta_{\mathcal P}^{t+1}(1)=&p_{\mathcal P}(1|1,\sigma (1)=CN)\eta_{\mathcal P}^{t}(1)\\&+p_P(1|0,\sigma (0)=CN)\eta_{\mathcal P}^{t}(0)\\
\end{array}
\right.
\end{equation}
Where $p_{\mathcal P}(\theta'|\theta,\sigma (\theta))$ denotes the transition probability that a $\theta$-worker from the current period becomes a $\theta'$-worker in the next period under the rating protocol $\mathcal P$ when the worker's action is $\sigma (\theta)$, which can be expressed as

$p_{\mathcal P}(\theta'|\theta,\sigma (\theta))=$
\begin{equation}{}
~~~~\left\{
\begin{array}{ll}
1-\beta(\varepsilon_1+\varepsilon_2-\varepsilon_1\varepsilon_2),&\emph{if}~ \sigma (\theta)=CN,\\&\theta=1 ~\emph{and}~ \theta'=1\\
\beta(\varepsilon_1+\varepsilon_2-\varepsilon_1\varepsilon_2),&\emph{if}~ \sigma (\theta)=CN,\\&\theta=1 ~\emph{and}~ \theta'=0\\
\alpha(1-\varepsilon_1-\varepsilon_2+\varepsilon_1\varepsilon_2),&\emph{if}~ \sigma (\theta)=CN,\\&\theta=0 ~\emph{and}~ \theta'=1\\
1-\alpha(1-\varepsilon_1-\varepsilon_2+\varepsilon_1\varepsilon_2),&\emph{if}~ \sigma (\theta)=CN,\\&\theta=0 ~\emph{and}~ \theta'=0\\
1-\beta(1-\varepsilon_2+\varepsilon_1\varepsilon_2),&\emph{if}~ \sigma (\theta)=CA,\\&\theta=1 ~\emph{and}~ \theta'=1\\
\beta(1-\varepsilon_2+\varepsilon_1\varepsilon_2),&\emph{if}~ \sigma (\theta)=CA,\\&\theta=1 ~\emph{and}~ \theta'=0\\
\alpha(\varepsilon_2-\varepsilon_1\varepsilon_2),&\emph{if}~ \sigma (\theta)=CA,\\&\theta=0 ~\emph{and}~ \theta'=1\\
1-\alpha(\varepsilon_2-\varepsilon_1\varepsilon_2),&\emph{if}~ \sigma (\theta)=CA,\\&\theta=0 ~\emph{and}~ \theta'=0\\
\end{array}
\right.
\end{equation}

As $\sum_{\theta=0}^1\eta_{\mathcal P}(\theta)=1$, with simple manipulations, there exists a unique stationary rating distribution $\{\eta_{\mathcal P}^t(\theta)\}$, which is derived as follows:
\begin{equation}{}
\label{distribution}
\left\{
\begin{array}{ll}
\eta_{\mathcal P}(0)=\frac{\beta(\varepsilon_1+\varepsilon_2-\varepsilon_1\varepsilon_2)} {\beta(\varepsilon_1+\varepsilon_2-\varepsilon_1\varepsilon_2)+\alpha(1-\varepsilon_1-\varepsilon_2+\varepsilon_1\varepsilon_2)}\\
\eta_{\mathcal P}(1)=\frac{\alpha(1-\varepsilon_1-\varepsilon_2+\varepsilon_1\varepsilon_2)} {\beta(\varepsilon_1+\varepsilon_2-\varepsilon_1\varepsilon_2)+\alpha(1-\varepsilon_1-\varepsilon_2+\varepsilon_1\varepsilon_2)}\\
\end{array}
\right.
\end{equation}

\subsection{Sustainable Conditions}
Given a fixed pricing scheme $\psi$ under the rating protocol $\mathcal P$, the expected one-period utility of a worker is only determined by his adopted strategy as well as his current rating label, which can be expressed as follows:
\begin{equation}{v_{\mathcal P}(\theta|\alpha)=}
\left\{
\begin{array}{ll}
v_{CN}(\gamma_1),&\emph{if} ~\theta=1 ~\emph{and}~ \sigma (\theta)=CN\\
v_{CN}(\gamma_0),&\emph{if} ~\theta=0 ~\emph{and}~ \sigma (\theta)=CN\\
v_{CA}(\gamma_1),&\emph{if} ~\theta=1 ~\emph{and}~ \sigma (\theta)=CA\\
v_{CA}(\gamma_0),&\emph{if} ~\theta=0 ~\emph{and}~ \sigma (\theta)=CA\\
\end{array}
\right.
\end{equation}
Where the expected payoff associated with actions \emph{CN} and \emph{CA} under monitoring or reporting errors can be found according to Eq.(\ref{payoff2}) in Appendix B. Since a non-compliant worker may benefit by unilaterally deviating from the recommended strategy \emph{CN}, and myopically choosing strategies \emph{CA}, \emph{SA} or \emph{SN}, then his expected payoff has its maximal value when it chooses strategy \emph{CA} with probability 1, \emph{i.e.}, a worker's optimal strategy choice is binary regardless of his current rating label: either \emph{CN} or \emph{CA}, and hence there is no need to consider the remaining two strategies \emph{SA} and \emph{SN} that a non-compliant worker may choose.

The expected long-term utility of a $\theta$-worker is the infinite-horizon discounted sum of his expected one-period utility with his expected future payoff multiplied by a common discount factor $\delta$, which can be computed by solving the following recursive equation
\begin{equation} {}
\label{eq5}
v_{\mathcal P}^\infty (\theta|\sigma (\theta))= v_{\mathcal P} (\theta|\sigma (\theta))+ \delta \sum_{\theta' \in \Theta} p_{\mathcal P} (\theta' |\theta,\sigma (\theta))v_{\mathcal P}^\infty (\theta'|\sigma (\theta))
\end{equation}
Where $\delta \in [0,1)$ is the rate at which a worker discounts his future payoff, and reflects his patience. It is obvious that a more patient worker has a larger discount factor.

According to the self-interested strategic nature of workers, they are willing to comply with the recommended strategy $\sigma$, if and only if they find it advantageous to their self-interest, \emph{i.e.}, they cannot be benefitted with respect to their expected long-term utility upon deviations given the deployed rating protocol $\mathcal P$. Such a rating protocol is called a sustainable rating protocol, and the sustainability of it is correspondingly defined as follows:

\begin{myDef}
A rating protocol $\mathcal P$ is sustainable if the optimal social strategy for a $\theta$-worker is $\sigma_{\mathcal P}^\ast(\theta)=CN,\forall\theta \in \Theta$.
\end{myDef}

When a $\theta$-worker follows the recommend strategy $\sigma_{\mathcal P}(\theta)=CN$ under the rating protocol $\mathcal P$, he receives expected long-term utility $v_{\mathcal P}^\infty(\theta|\sigma_{\mathcal P}(\theta))$. On the contrary, he will receive expected long-term utility $v_{\mathcal P}^\infty(\theta|\sigma_{\mathcal P}'(\theta))$ when he deviates from the recommended strategy $CN$ to $CA$. By comparing these two payoffs, we can derive necessary and sufficient conditions for a rating protocol to be sustainable, as shown in Proposition 1.

\newtheorem{proposition}{Proposition}

\begin{proposition}
 A rating protocol $\mathcal P$ is sustainable if and only if
 \begin{equation} {}
 \label{eq6}
   \begin{split}
   &v_{\mathcal P}^\infty (1|\sigma^\ast)-v_{\mathcal P}^\infty (0|\sigma^\ast)\ge\\
   &\frac {max\Big\{\big\lbrack v_{CA}(\gamma_0)-v_{CN}(\gamma_0)\big\rbrack/\alpha,[v_{CA}(\gamma_1)-v_{CN}(\gamma_1)]/\beta\Big\}} {\delta(1-\varepsilon_1-2\varepsilon_2+2\varepsilon_1\varepsilon_2)}
   \end{split}
   \end{equation}
\end{proposition}

\begin{IEEEproof}
For the ``if'' part: Given a sustainable rating protocol $\mathcal P$, a $\theta$-worker will comply with the optimal social strategy with $\sigma_{\mathcal P}^\ast(0)=\sigma_{\mathcal P}^\ast(1)=CN$, his expected long-term utilities can be expressed as follows:
 \begin{equation} {}
 \label{eq7}
   \begin{split}
   v_{\mathcal P}^\infty (0|\sigma^\ast)=v_{CN} (\gamma_0)+ \delta\big\lbrack p_{\mathcal P} (1|0,\sigma^\ast)v_{\mathcal P}^\infty (1|\sigma^\ast)+\\p_{\mathcal P} (0|0,\sigma^\ast)v_{\mathcal P}^\infty (0|\sigma^\ast)\big\rbrack
   \end{split}
   \end{equation}

   \begin{equation} {}
   \label{eq8}
   \begin{split}
   v_{\mathcal P}^\infty (1|\sigma^\ast)=v_{CN} (\gamma_1)+ \delta\big\lbrack p_{\mathcal P} (1|1,\sigma^\ast)v_{\mathcal P}^\infty (1|\sigma^\ast)+\\p_{\mathcal P} (0|1,\sigma^\ast)v_{\mathcal P}^\infty (0|\sigma^\ast)\big\rbrack
   \end{split}
   \end{equation}
By comparing (\ref{eq7}) and (\ref{eq8}), we have
\begin{equation} {}
\label{eq9}
   \begin{split}
   &v_{\mathcal P}^\infty (1|\sigma^\ast)-v_{\mathcal P}^\infty (0|\sigma^\ast)=\\
   &\frac {v_{CN}(\gamma_1)-v_{CN}(\gamma_0)} {1-\delta\big\lbrack 1-\beta(\varepsilon_1+\varepsilon_2-\varepsilon_1\varepsilon_2)-\alpha(1-\varepsilon_1-\varepsilon_2+\varepsilon_1\varepsilon_2)\big\rbrack }
   \end{split}
   \end{equation}
According to the one-shot deviation principle \cite{29}, the expected long term utility of a worker with rating 0 unilaterally deviating from $\sigma^\ast$ to $\sigma'$ only in the current period and following $\sigma^\ast$ afterwards, which can be computed by solving
\begin{equation} {}
\label{eq10}
   \begin{split}
   v_{\mathcal P}^\infty (0|\sigma')=v_{CA} (\gamma_0)+ \delta\big\lbrack p_{\mathcal P} (1|0,\sigma' (0)=CA)v_{\mathcal P}^\infty (1|\sigma^\ast)+\\p_{\mathcal P} (0|0,\sigma' (0)=CA)v_{\mathcal P}^\infty (0|\sigma^\ast)\big\rbrack
   \end{split}
   \end{equation}
Similarly, the expected long-term utility of a deviating worker with rating 1 is given by
\begin{equation} {}
\label{eq11}
   \begin{split}
   v_{\mathcal P}^\infty (1|\sigma')=v_{CA} (\gamma_1)+ \delta\big\lbrack p_{\mathcal P} (1|1,\sigma' (1)=CA)v_{\mathcal P}^\infty (1|\sigma^\ast)+\\p_{\mathcal P} (0|1,\sigma' (1)=CA)v_{\mathcal P}^\infty (0|\sigma^\ast)\big\rbrack
   \end{split}
   \end{equation}
By solving inequalities (\ref{eq12}) and (\ref{eq13})
\begin{equation} {}
\label{eq12}
v_{\mathcal P}^\infty (0|\sigma^\ast)\ge v_{\mathcal P}^\infty (0|\sigma')
\end{equation}
\begin{equation} {}
\label{eq13}
v_{\mathcal P}^\infty (1|\sigma^\ast)\ge v_{\mathcal P}^\infty (1|\sigma')
\end{equation}
we can obtain the following inequality by substituting (\ref{eq7}) into the LHS of (\ref{eq12}), and substituting (\ref{eq10}) into RHS of (\ref{eq12}).
\begin{equation} {}
\label{eq14}
v_{\mathcal P}^\infty (1|\sigma^\ast)-v_{\mathcal P}^\infty (0|\sigma^\ast)\ge
   \frac {v_{CA}(\gamma_0)-v_{CN}(\gamma_0)} {\delta\alpha(1-\varepsilon_1-2\varepsilon_2+2\varepsilon_1\varepsilon_2)}
\end{equation}	
Similarly, the following inequality can be derived by substituting (\ref{eq8}) into the LHS of (\ref{eq13}), and substituting (\ref{eq11}) into RHS of (\ref{eq13}).
\begin{equation} {}
\label{eq15}
v_{\mathcal P}^\infty (1|\sigma^\ast)-v_{\mathcal P}^\infty (0|\sigma^\ast)\ge
   \frac {v_{CA}(\gamma_1)-v_{CN}(\gamma_1)} {\delta\beta(1-\varepsilon_1-2\varepsilon_2+2\varepsilon_1\varepsilon_2)}
\end{equation}
Hence, inequality (\ref{eq6}) can be hold by combing inequalities (\ref{eq14}) and (\ref{eq15}).

For the ``only if'' part: Suppose that inequality (\ref{eq6}) is satisfied under the rating protocol $\mathcal P$, we can obtain $v_P^\infty(0|\sigma^\ast)\ge v_P^\infty(0|\sigma)$ and $v_P^\infty(1|\sigma^\ast)\ge v_P^\infty(1|\sigma)$ for any $\sigma$. With this in mind, we can derive the range for these design parameters (\emph{i.e.}, $\alpha$, $\beta$, $\gamma_1$ and $\gamma_0$) in (\ref{eq6}), and thus design a rating protocol under which sustains the incentive for workers to comply with the recommended strategies. Hence, this proposition follows.
\end{IEEEproof}

\subsection{Rating Protocol Design with Constraints}
Given a sustainable rating protocol $\mathcal P$, each worker always chooses to devote a high level of effort in the first stage and does not attack his opponent in the second stage, hence, the requester's utility will be only determined by the payment that he rewards the winner, which is $\gamma_1$ to the winner having rating 1, and $\gamma_0$ to the winner having rating 0. Let $u_{\mathcal P}^\theta$  denote the expected one-period utility of the requester in the case that the winner has rating $\theta$, taking into account monitoring or reporting errors $\varepsilon_1$ and $\varepsilon_2$. The expression of $u_{\mathcal P}^\theta$ can be derived as follows:
\begin{equation} {}
u_{\mathcal P}^\theta=(1-\varepsilon_1-\varepsilon_2+\varepsilon_1\varepsilon_2)(1-\gamma_\theta)+(\varepsilon_1+\varepsilon_2-\varepsilon_1\varepsilon_2)(-\gamma_\theta)
\end{equation}
Where $1-\gamma_\theta$ is the utility of the requester when he reports that his task has been fulfilled, which happens with probability $1-\varepsilon_1-\varepsilon_2+\varepsilon_1\varepsilon_2$. When the task has not been accomplished, which happens with probability $\varepsilon_1+\varepsilon_2-\varepsilon_1\varepsilon_2$, the utility of the requester is $-\gamma_\theta$, \emph{i.e.}, the request suffers a payment $\gamma_\theta$, and receives no benefit.

In a two-stage game of a crowdsourcing competition, the worker who has a higher productivity at the end of the second stage wins reward $\gamma_\theta$, whose value is determined by the winner's rating label $\theta$.  Let $(\theta,\tilde{\theta})$ denote a focal worker with rating $\theta$ competing against his opponent worker with rating $\tilde{\theta}$, to win the reward. There are 4 possible combinations for $(\theta,\tilde{\theta})$ namely (0,0), (0,1), (1,0) and (1,1). The computation process for the social utility of the requester in these four cases is as follows:
\begin{equation} {}
\label{eq28}
   \begin{split}
   &u_{\mathcal P}(\theta,\tilde{\theta})=\eta(\theta)\eta(\theta_0)(\frac{1}{2}u_{\mathcal P}^{\theta}+\frac{1}{2}u_{\mathcal P}^{\tilde{\theta}})\\
   &=\left\{
\begin{array}{ll}
\eta(0)\eta(0)u_{\mathcal P}^0,&\emph{if}~ \theta=\tilde{\theta}=0\\
\frac{1}{2}\eta(0)\eta(1)(u_{\mathcal P}^0+u_{\mathcal P}^1),&\emph{if}~ \theta=0 ~\emph{and}~ \tilde{\theta}=1\\
\frac{1}{2}\eta(0)\eta(1)(u_{\mathcal P}^0+u_{\mathcal P}^1),&\emph{if}~ \theta=1 ~\emph{and}~ \tilde{\theta}=0\\
\eta(1)\eta(1)u_{\mathcal P}^1,&\emph{if}~ \theta=\tilde{\theta}=1\\
\end{array}
\right.
   \end{split}
   \end{equation}

$\eta(\theta)\eta(\theta_0)$ is the probability that the focal worker with a rating $\theta$ competes against the opponent worker whose rating is $\tilde{\theta}$. When both of the focal worker and his opponent worker comply with the recommended strategy, each one is equally likely to win the reward, and the expected probability for each one to be the winner is $1/2$. Given a sustainable rating protocol $\mathcal P$, strategic workers will find enticement to follow the recommended strategy in their self-interest regardless of their ratings even under the imperfect monitoring, and which worker wins the game depends only on the strategies adopted by both workers, hence, the probability of becoming the winner is independent of the workers' ratings and thus each worker has the same probability of becoming the winner.

Let $U_{\mathcal P}$ denote the social utility under the sustainable protocol $\mathcal P$, which can be formulated as follows:
\begin{equation} {}
\label{eq27}
U_{\mathcal P}=\sum_{\theta=0}^1\sum_{\tilde{\theta}=0}^1u_{\mathcal P}(\theta,\tilde{\theta})
\end{equation}

Eq.(\ref{eq27}) can be rewritten as follows by substituy Eq.(\ref{eq28}) to it:
\begin{equation} {}
\label{eq280}
   \begin{split}
U_{\mathcal P}=&(1-\varepsilon_1-\varepsilon_2+\varepsilon_1\varepsilon_2)-\\
&\frac{\beta\gamma_0(\varepsilon_1+\varepsilon_2-\varepsilon_1\varepsilon_2)+\alpha\gamma_1(1-\varepsilon_1-\varepsilon_2+\varepsilon_1\varepsilon_2)} {\beta(\varepsilon_1+\varepsilon_2-\varepsilon_1\varepsilon_2)+\alpha(1-\varepsilon_1-\varepsilon_2+\varepsilon_1\varepsilon_2)}
   \end{split}
   \end{equation}

We assume that the protocol designer is profit-seeking and aims to design a sustainable rating protocol $\mathcal P$ that maximizes the requester's expected one-period utility, which is the benefit per period of a requester, deducts the payment reward to workers. In order to attract workers to stay in the platform for a long period of time, additional incentive constraints are needed to prevent the long period utility of a worker to be negative, regardless of his current rating label. Given this, the problem of designing a rating protocol can be formally expressed as:

\begin{myDef}
The rating protocol design problem is formulated as follows:
\begin{equation}{}
\label{eq29}
\left\{
\begin{array}{ll}
&\mathop{\max}\limits_{(\alpha,\beta,\gamma_1,\gamma_0)}U_{\mathcal P}\triangleq \sum_{\theta=0}^1\sum_{\tilde{\theta}=0}^1u_{\mathcal P}(\theta,\tilde{\theta})\\
&s.t.
\left\{
\begin{array}{ll}\sigma_{\mathcal P}^\ast(\theta)=CN,&\forall\theta\in\Theta\\u_{\mathcal P}^\infty(\theta)\ge 0,&\forall\theta\in\Theta \end{array}\right.\\
\end{array}
\right.
\end{equation}
\end{myDef}

\begin{figure*}
\begin{equation}{}
\label{eq30}
\left\{
\begin{array}{ll}
K_1^i=\frac{(1-\varepsilon_1-2\varepsilon_2+2\varepsilon_1\varepsilon_2)\big\lbrack v_{CN}^i(\gamma_1)-v_{CN}^i(0)\big\rbrack -(1-\varepsilon_1-\varepsilon_2+\varepsilon_1\varepsilon_2)\big\lbrack v_{CA}^i(0)-v_{CN}^i(0)\big\rbrack } {(\varepsilon_1+\varepsilon_2-\varepsilon_1\varepsilon_2)\big\lbrack v_{CA}^i(0)-v_{CN}^i(0)\big\rbrack }\\
B_1^i=-\frac{1-\delta}{\delta(\varepsilon_1+\varepsilon_2-\varepsilon_1\varepsilon_2)}\\
K_2^i=\frac{(1-\varepsilon_1-\varepsilon_2+\varepsilon_1\varepsilon_2)\big\lbrack v_{CA}^i(\gamma_1)-v_{CN}^i(\gamma_1)\big\rbrack } {(1-\varepsilon_1-2\varepsilon_2+2\varepsilon_1\varepsilon_2)\big\lbrack v_{CN}^i(\gamma_1)-v_{CN}^i(0)\big\rbrack -(\varepsilon_1+\varepsilon_2-\varepsilon_1\varepsilon_2)\big\lbrack v_{CA}^i(\gamma_1)-v_{CN}^i(\gamma_1)\big\rbrack }\\
B_2^i=\frac{(1-\delta)\big\lbrack v_{CA}^i(\gamma_1)-v_{CN}^i(\gamma_1)\big\rbrack } {\delta(1-\varepsilon_1-2\varepsilon_2+2\varepsilon_1\varepsilon_2)\big\lbrack v_{CN}^i(\gamma_1)-v_{CN}^i(0)\big\rbrack -\delta(\varepsilon_1+\varepsilon_2-\varepsilon_1\varepsilon_2)\big\lbrack v_{CA}^i(\gamma_1)-v_{CN}^i(\gamma_1)\big\rbrack }\\
K_3^i=-\frac{(1-\varepsilon_1-\varepsilon_2+\varepsilon_1\varepsilon_2)v_{CN}^i(\gamma_1)} {(\varepsilon_1+\varepsilon_2-\varepsilon_1\varepsilon_2)v_{CN}^i(0)}\\
B_3^i=\frac{\delta-1} {\delta(\varepsilon_1+\varepsilon_2-\varepsilon_1\varepsilon_2)}\\
K_4=\frac{(1-\varepsilon_1-\varepsilon_2+\varepsilon_1\varepsilon_2)\big\lbrack (1-\varepsilon_1-\varepsilon_2+\varepsilon_1\varepsilon_2)-\gamma_1-U_{\mathcal P}\big\rbrack } {(\varepsilon_1+\varepsilon_2-\varepsilon_1\varepsilon_2)\big\lbrack U_{\mathcal P}-(1-\varepsilon_1-\varepsilon_2+\varepsilon_1\varepsilon_2)\big\rbrack }\\
\end{array}
\right.
\end{equation}
\end{figure*}

\section{OPTIMAL DESIGN OF RATING PROTOCOLS}

In this section we investigate the design of an optimal rating protocol that solves the design problem of Eq. (\ref{eq29}), \emph{i.e.}, selecting the optimal rating scheme $\tau$ and optimal pricing scheme $\psi$, which are determined by four design parameters $(\alpha,\beta,\gamma_1,\gamma_0)$.  In order to characterize an optimal design, which is denoted as $(\alpha^\ast,\beta^\ast,\gamma_1^\ast,\gamma_0^\ast)$, we investigate the impacts of four design parameters on the social welfare $U_{\mathcal P}=\sum_{\theta=0}^1\sum_{\tilde{\theta}=0}^1u_{\mathcal P}(\theta,\tilde{\theta})$, and the incentive for satisfying constraints in Eq. (\ref{eq29}).

\subsection{Constraints in the Optimal Design Problem}
In order to maximize the social utility, the requester wants to pay his workers as little as possible, on the premise of supporting enough incentive for workers to follow the recommend strategy \emph{CN}, and stay in the system for a long period. The above observations are summarized in the following theorem.

\begin{myTheo}
$\gamma_0^\ast=0$ is always an optimal solution to (\ref{eq29}).
\end{myTheo}
\begin{IEEEproof}
Given $\alpha,\beta$ and $\gamma_1$, the social utility monotonically decreases with $\gamma_0$, which is  maximized when $\gamma_0=0$. Let us assume that a worker deviates from the social norm, his rating will be decreased to 0 with a large probability, and he will receive the reward $\gamma_0$ as a punishment. Hence, the worker's incentive is also maximized when  $\gamma_0=0$. Therefore, this statement follows.
\end{IEEEproof}

In the remainder of our design, we set $\gamma_0=0$ by default without further notice. Given a fixed $\gamma_1$, variables $K_1^i$, $B_1^i$, $K_2^i$, $B_2^i$, $K_3^i$, $B_3^i$ and $K_4$ are defined in Eq. (\ref{eq30}) for convenience.

 \begin{proposition}
 Given a fixed $\gamma_1$, constraints in the rating protocol design problem (\ref{eq29}) for $i$th-worker are satisfied if and only if $\beta\in[K_2^i\alpha+B_2^i,K_3^i\alpha+B_3^i]$.

\end{proposition}
\begin{IEEEproof}
For the ``if'' part: Given a fixed $\gamma_1$, we assume that each constraint in the rating protocol design problem (\ref{eq29}) are satisfied, there are two constraints that need to be fulfilled simultaneously for the \emph{i}th-worker.

Firstly, we need to satisfy the one-shot deviation principle, and thus have the lower bound of $\beta$ by substituting (\ref{eq9}) into (\ref{eq6}), \emph{i.e.}, $\beta\in[\max\{K_1^i\alpha+B_1^i,K_2^i\alpha+B_2^i\},1]$. If the \emph{i}th-worker holds rating 0, then his expected one-period utility $v_{\mathcal P}^i(\gamma_0^\ast)<0$, this is because the \emph{i}th-worker will receive no reward but consume $c_i$ by choosing \emph{C} in the first stage. The expected one-period utility of a worker with rating 0 whose adopted strategy is \emph{CA} will loss more than the worker with rating 0 whose adopted strategy is \emph{CN}, because the former one consumes $c_i$ in the first stage, and $s_i$ in the second stage, but the latter one only consumes $c_i$ in the first stage, \emph{i.e.}, $v_{CA}^i(0)<v_{CN}^i(0)<0$. Therefore, we have $K_1^i<0$, $B_1^i<0$ and $\beta\ge K_1^i\alpha+B_1^i$ regardless of any choice of $\alpha$ and $\beta$, we thus have $\beta\in[K_2^i\alpha+B_2^i,1]$.

Secondly, in order to make sure that workers have sufficient incentive to contribute good behaviors, we must hold both $v_{\mathcal P}^\infty(0)\ge0$ and $v_{\mathcal P}^\infty(1)\ge0$. Since the worker with rating 1 will obtain reward $\gamma_1$, which is higher than the worker with rating 0 whose reward is $\gamma_0$. We thus only have to make sure that the long-term utility of the worker with rating zero must be non-negative which corresponds to the upper bound of $\beta$ by substituting (\ref{eq9}) into (\ref{eq7}), \emph{i.e.}, $\beta\in [0,K_3^i\alpha+B_3^i]$. Hence, this statement follows.

For the ``only if'' part, we assume that $\beta\in[K_2^i\alpha+B_2^i,K_3^i\alpha+B_3^i]$, it is easy to determine that constraints in the rating protocol design problem (\ref{eq29}) are satisfied, and hence the ``only if'' part can be proved, which is omitted here.
\end{IEEEproof}

 \begin{proposition}
Given a fixed $\gamma_1$, we have
\begin{enumerate}[(i)]
  \item     $\beta=K_4\alpha$;		
  \item 	Social utility monotonically increases with $K_4$.
\end{enumerate}
\end{proposition}
\begin{IEEEproof}
(\emph{i}) $\beta=K_4\alpha$ is directly converted from (\ref{eq280}), in which we treat $K_4$ as a constant and $\beta$ as a function with variable $\alpha$.

(\emph{ii}) The partial derivative $U_{\mathcal P}$ with respect to $K_4$ can be derived as follows.
\begin{equation} {}
\label{eq34}
\frac{\partial U_{\mathcal P}} {\partial K_4}=\frac{(\varepsilon_1+\varepsilon_2-\varepsilon_1\varepsilon_2)\big\lbrack U_{\mathcal P}-(1-\varepsilon_1-\varepsilon_2+\varepsilon_2\varepsilon_2)\big\rbrack ^2} {(1-\varepsilon_1-\varepsilon_2+\varepsilon_1\varepsilon_2)\gamma_1}
\end{equation}

It is obvious that the value of (\ref{eq34}) is positive, hence the statement (\emph{ii}) follows.
\end{IEEEproof}

\subsection{Optimal Value of the Design Problem}

We now focus on the optimal value of the remaining three design parameters $\alpha,\beta$ and $\gamma_1$. As shown in the proof of Proposition 2, there exist two constraints that need to be fulfilled simultaneously for the \emph{i}th-worker. Similary, there exist another two constraints for the \emph{j}th-worker, and hence we have $\beta\in[\max\{K_2^i\alpha+B_2^i,K_2^j\alpha+B_2^j\},\min\{K_3^i\alpha+B_3^i,K_3^j\alpha+B_3^j\}]$

In the following, we design a rating protocol that achieves the social optimum at the equilibrium under the conditions that all of these four constraints hold. We assume that $\gamma_1$ is fixed, the four constraints constitute a convex set \emph{Q}, \emph{i.e.}, if and only if $(\alpha,\beta)\in Q$, the two constraints in Proposition 2 will be satisfied simultaneously. Since the feasible domain of $(\alpha,\beta)$ is a convex set, the optimal solution $(\alpha^\ast,\beta^\ast)$ must be at the boundary of the feasible domain, \emph{i.e.}, $\alpha^\ast=1$ or $\beta^\ast=1$, thereby we find the local optimum in two cases and then find a global optimum from such two cases.

\begin{algorithm}[bt]
\caption{Optimal Design of Rating Protocol}
\label{alg1}
\begin{algorithmic}[1]
\REQUIRE $c_1$, $c_2$, $s_1$, $s_2$, $\varepsilon_1$, $\varepsilon_2$, $d$ and $\delta$;
\ENSURE $\alpha^\ast$, $\beta^\ast$, $\gamma_1^\ast$;

\STATE $K_2=\max_{i\in\{1,2\}}K_3^i$;
\STATE $B_2=B_2^{\arg_{i\in\{1,2\}}\max K_2^i}$;
\STATE $K_3=\min_{i\in\{1,2\}}K_3^i$;
\STATE $B_3=B_3^{\arg_{i\in\{1,2\}}\min K_3^i}$;

\IF {$\gamma_1\in[{\underline{\gamma_{1,1}}},{\overline{\gamma_{1,1}}}]$ satisfies $(\frac{1-B_3}{K_3}<1)\cap \Big\{(K_2<0)\cup\big\lbrack(K_2>0)\cap(\frac{1-B_2}{K2}>\frac{1-B_3}{K_3})\big\rbrack\Big\}$}
\STATE $\alpha_1^\ast=\frac{1-B_3}{K_3}$
\STATE $\beta_1^\ast=1$
\STATE $\gamma_{1,1}^\ast={\underline{\gamma_{1,1}}}$
\STATE {Compute }$U_{\mathcal P}^1(\alpha_1^\ast,\beta_1^\ast,\gamma_{1,1}^\ast)$
\ENDIF

\IF {${\gamma_2\in[\underline{\gamma_{1,2}}},{\overline{\gamma_{1,2}}}]$ satisfies $\gamma_1$ in $(K_3+B_3\le1)\cap\Big\{(K_2<0)\cup\big\lbrack(K_2>0)\cap(K_2+B_2\le K_3+B_3)\big\rbrack\Big\}$}
\STATE $\alpha_2^\ast=1$
\STATE $\beta_2^\ast=K_3+B_3$
\STATE $\gamma_{1,2}^\ast={\overline{\gamma_{1,2}}}$
\STATE {Compute }$U_{\mathcal P}^2(\alpha_2^\ast,\beta_2^\ast,\gamma_{1,2}^\ast)$
\ENDIF
\STATE $t=\mathop{\arg}\limits_{i\in\{1,2\}}\max~U_{\mathcal P,i}$
\STATE $\alpha^\ast=\alpha_t^\ast$
\STATE $\beta^\ast=\beta_t^\ast$
\STATE $\gamma_1^\ast=\gamma_{1,t}^\ast$
\end{algorithmic}
\end{algorithm}

\begin{figure*}
\begin{minipage}{0.24\textwidth}
\centering
\includegraphics[height=4cm,width=4cm,angle=0,scale=1]{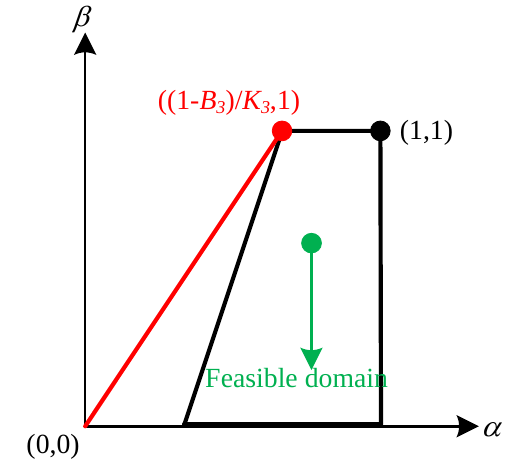}
\centerline{(\emph{a}) Case $\romannumeral1$: $\beta^\ast=1\wedge K_2<0$}
\end{minipage}
\begin{minipage}{0.24\textwidth}
\centering
\includegraphics[height=4cm,width=4cm,angle=0,scale=1]{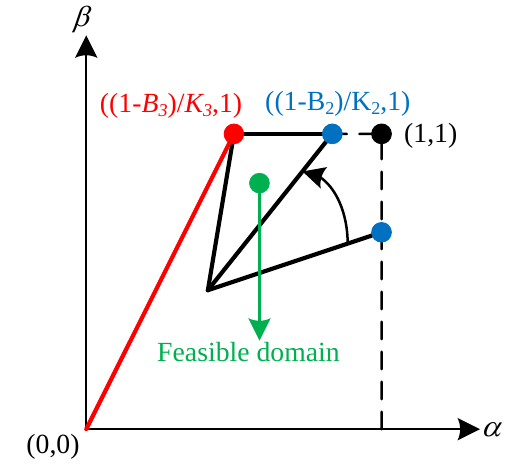}
\centerline{(\emph{b}) Case $\romannumeral2$: $\beta^\ast=1\wedge K_2>0$}
\end{minipage}
\begin{minipage}{0.24\textwidth}
\centering
\includegraphics[height=4cm,width=4cm,angle=0,scale=1]{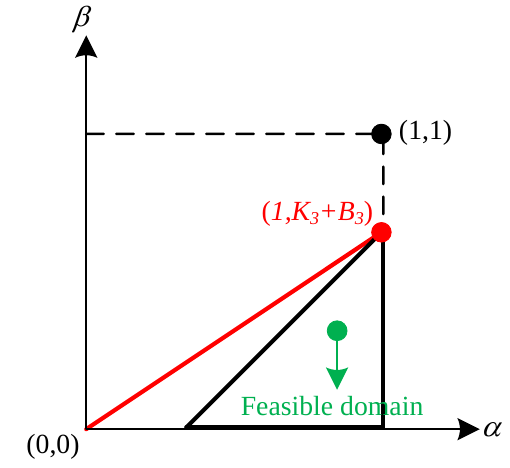}
\centerline{(\emph{c}) Case $\romannumeral3$: $\alpha^\ast=1\wedge K_2<0$}
\end{minipage}
\begin{minipage}{0.24\textwidth}
\centering
\includegraphics[height=4cm,width=4cm,angle=0,scale=1]{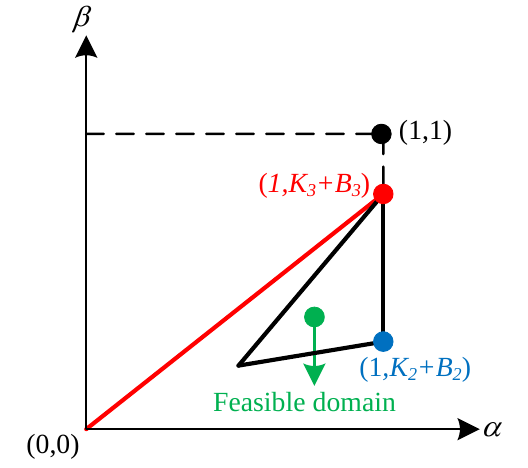}
\centerline{(\emph{d}) Case $\romannumeral4$: $\alpha^\ast=1\wedge K_2>0$}
\end{minipage}
 \caption{ Illustration of how to build the feasible domain for ($\alpha,\beta$): According to the possible optimal solution for $\alpha$ and $\beta$, as well as the range of $\gamma_1$, we divide such a problem into four cases, the closed region in each case is a feasible solution of $(\alpha,\beta)$ for a fixed $\gamma_1$, and the red point denotes the optimal solution $(\alpha^\ast,\beta^\ast)$.}
\label{fig:res}
\end{figure*}

\begin{myTheo}
The output $(\alpha^\ast,\beta^\ast,\gamma_1^\ast)$ derived by Algorithm 1 is an optimal solution of (\ref{eq29}).
\end{myTheo}
\begin{IEEEproof}
Algorithm \ref{alg1} takes $c_1$, $c_2$, $s_1$, $s_2$, $\varepsilon_1$, $\varepsilon_2$, $d$ and $\delta$ as input and returns $(\alpha^\ast,\beta^\ast,\gamma_1^\ast)$. It consists of 3 parts: Part 1 (line 1-4) determines the constrained parameters, \emph{i.e.}, $K_2$, $B_2$, $K_3$, and $B_3$; Part 2 (line 5-16) determines whether there exists a feasible solution with the remaining three design parameters satisfying constraints in (\ref{eq29}). If it is true, then we obtain a partial optimal solution; Part 3 (line 17-20) returns the global optimal solution $(\alpha^\ast,\beta^\ast,\gamma_1^\ast)$ according to all the local optimal solutions we have found. The computational complexity of Algorithm 1 is $O(mn^2)$, where \emph{m} and \emph{n} represent the reciprocal of accumulation unit for determining $\gamma_1^\ast$ and the number of iterations for computing the expected long-term utility in (\ref{eq5}), respectively. The detailed explanation of this algorithm is as follows:

Part 1 (line 1-4): According to (\ref{eq30}), it is easy to determine that $B_2^i>B_2^j$ if $K_2^i>K_i^j$, and we thus have $\beta\ge K_2^i\alpha+B_2^i>K_2^j\alpha+B_2^j$. If we require that these two constraints hold simultaneously, we only need to make sure $\beta\ge K_2^i\alpha+B_2^i$, as shown in line 1 and 2. Similarly, $\beta\le K_3^j\alpha+B_3^j<K_3^i\alpha+B_3^i$ if $K_3^i>K_3^j$, we then only need to make sure $\beta\le K_3^j\alpha+B_3^j$, as shown in line 3 and 4.

Part 2 (line 5-16): It is obvious that the feasible set $Q$ formed by constraints $\beta\in[K_2\alpha+B_2,K_3\alpha+B_3]\cap[0,1]$ and $\alpha\in[0,1]$ is a convex set, and hence the optimal values of $\alpha$ and $\beta$ must at the endpoint. According to (\ref{eq30}), it is easy to determine that $K_3^i>0$, $B_3^i<0$, and both $K_2^i$ and $B_2^i$ are positive or negative. According to Proposition 3, $\beta=K_4\alpha$ and the social utility monotonically increases with $K_4$. Therefore, we obtain two possible solutions that either $\alpha^\ast=1$ or $\beta^\ast=1$, and divide it into four cases as shown in Figure 2 by graphical method for the linear programming problem.

Case $\romannumeral1$ and $\romannumeral2$ (line 5-10): The feasible domain of $(\alpha,\beta)$ is illustrated as shown in Figure 2(a) and 2(b) when $\beta^\ast=1$. We assume that $\beta_1^\ast=1\wedge K_2<0$ for Case $\romannumeral1$, and have $B_2<0$ according to (\ref{eq30}). Therefore, it is easy to determine that $\beta\ge K_2\alpha+B_2,\forall \alpha, \beta\in[0,1]$ holds. Moreover, we need to make sure that $\frac{1-B_3} {K_3}<1$ holds for $\{\beta\le K_3\alpha+B_3\}\wedge \{\beta_1^\ast=1\}$. In Case $\romannumeral2$ as shown in Figure 2(b), we assume that $\beta_1^\ast=1\wedge K_2>0$, and ensure to hold $\frac{1-B_2}{K_2}>\frac{1-B_3}{K_3}$ for $\beta\in[K_2\alpha+B_2,K_3\alpha+B_3]$. As the social utility increases as the slope increases, $(\alpha_1^\ast,\beta_1^\ast)=(\frac{1-B_3}{K3},1)$ is directly derived. Let $\underline{\gamma_{1,1}}$ and $\overline{\gamma_{1,1}}$ denote the smallest and largest value of $\gamma_{1,1}$ (\emph{i.e.}, the value of $\gamma_{1,1}$ when $\beta^\ast=1$ is fixed), respectively, such that $Q_1=(\frac{1-B_3}{K_3}<1)\cap \{(K_2<0)\cup[(K_2>0)\cap(\frac{1-B_2}{K2}>\frac{1-B_3}{K_3})]\}$ holds. Since the social utility decreases with $\gamma_{1,1}$, hence $\gamma_{1,1}^\ast={\underline{\gamma_{1,1}}}$. The expression of the social utility under the condition that $(\alpha_1^\ast , \beta_1^\ast)=(\frac{1-B_3}{K_3},1)$ holds is as follows (where we set $\mathcal Z=1-\varepsilon_1-\varepsilon_2+\varepsilon_1\varepsilon_2$):
\begin{equation} {}
   \begin{split}
   &U_{\mathcal P}^1(\alpha_1^\ast,\beta_1^\ast,\gamma_1)=\mathcal Z-\\
   &\frac{\gamma_1(1-\delta \mathcal Z)v_{CN}(\gamma_0)}{(1-\delta)v_{CN}(\gamma_0)+\delta(\varepsilon_1+\varepsilon_2-\varepsilon_1\varepsilon_2)\big\lbrack v_{CN}(\gamma_0)-v_{CN}(\gamma_1)\big\rbrack }
   \end{split}
   \end{equation}

Case $\romannumeral3$ and $\romannumeral4$ (line 11-16): The feasible domain of $(\alpha,\beta)$ is illustrated as shown in Figure 2(c) and 2(d) when $\alpha^\ast=1$. We assume that $K_2<0\wedge \alpha^\ast=1$ for Case $\romannumeral3$, which is similar with Case $\romannumeral1$, we only need to ensure $K_3+B_3\le1$ holds for $\{\beta\le K_3\alpha+B_3\}\wedge \{\alpha_1^\ast=1\}$. In Case $\romannumeral4$ as shown in Figure 2(d), we assume that $K_2>0\wedge \alpha^\ast=1$, and need to make sure that $\{K_3+B_3\le1\}\wedge \{K_2+B_2\le K_3+B_3\}$ holds for $\{\beta\in[K_2\alpha+B_2,K_3\alpha+B_3]\}\wedge \{\alpha^\ast=1\}$. Since the social utility monotonically increases as the slope increases, we then have $(\alpha_1^\ast,\beta_1^\ast)=(1,K_3+B_3)$. Let $\underline{\gamma_{1,2}}$ and $\overline{\gamma_{1,2}}$ denote the smallest and largest value of $\gamma_1$ when $\alpha^\ast=1$ is fixed, respectively, such that $Q_2=(K_3+B_3\le1)\cap\{(K_2<0)\cup[(K_2>0)\cap(K_2+B_2\le K_3+B_3)]\}$ holds. As the social utility increases with $\gamma_{1,2}$ when $(\alpha_1^\ast , \beta_1^\ast)=(1,K_3+B_3)$, we thus have $\gamma_{1,2}^\ast={\overline{\gamma_{1,2}}}$. The expression of the social utility under the condition that $(\alpha_1^\ast , \beta_1^\ast)=(1,K_3+B_3)$ is as follows:
\begin{equation} {}
   \begin{split}
   &U_{\mathcal P}^2(\alpha_2^\ast,\beta_2^\ast,\gamma_1)=\mathcal Z-\\
   &\frac{\delta \gamma_1 \mathcal Zv_{CN}(\gamma_0)}{(\delta-1)v_{CN}(\gamma_0)+\delta \mathcal Z\big\lbrack v_{CN}(\gamma_0)-v_{CN}(\gamma_1)\big\rbrack }
   \end{split}
   \end{equation}

In summary, all possible values of $\alpha$, $\beta$ and $\gamma$ are covered by the above four cases, and hence $\max\{U_{\mathcal P}^1,U_{\mathcal P}^2\}$ returns the optimal value of the social utility under the proposed rating protocol $\mathcal P$.
\end{IEEEproof}

\section{SIMULATION RESULTS}

In this section, we provide numerical results to illustrate the key features of the designed optimal rating protocol for the crowdsourcing contest dilemma. Firstly, we show how intrinsic parameters impact on design parameters. Secondly, we show how the performance gain of the proposed rating protocol changes when intrinsic parameters vary, especially under heterogeneity of workers. To solve problem (\ref{eq29}) of social welfare maximization, we first fix $\gamma_0^\ast=0$ regardless of any choice of intrinsic parameters, and then select $\alpha^\ast$, $\beta^\ast$ and $\gamma_1^\ast$ based on Algorithm 1.

\subsection{The Impact of Intrinsic Parameters on Design Parameters}

\begin{figure*}
\centering

\begin{minipage}{0.245\textwidth}
  \rightline{\includegraphics[width=\textwidth]{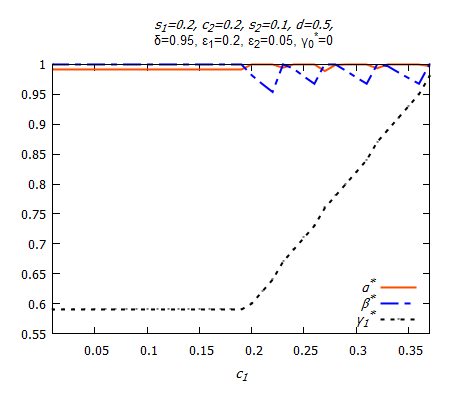}}
  \centerline{(\emph{a}) }
\end{minipage}
\begin{minipage}{0.245\textwidth}
  \leftline{\includegraphics[width=\textwidth]{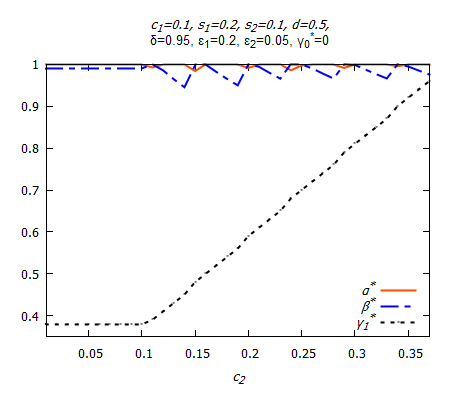}}
  \centerline{(\emph{b})}
\end{minipage}
\begin{minipage}{0.245\textwidth}
  \rightline{\includegraphics[width=\textwidth]{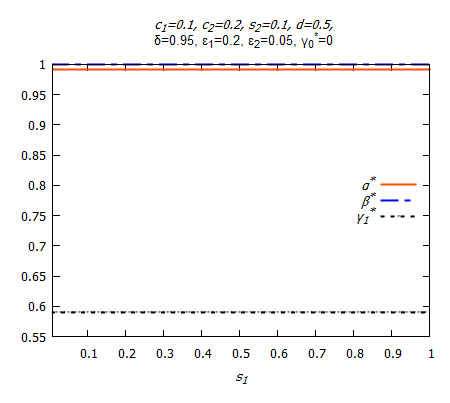}}
  \centerline{(\emph{c}) }
\end{minipage}
\begin{minipage}{0.245\textwidth}
  \leftline{\includegraphics[width=\textwidth]{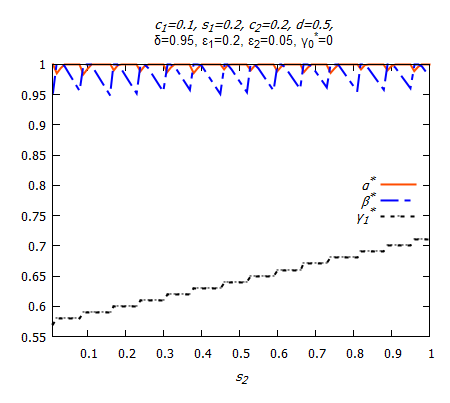}}
  \centerline{(\emph{d})}
\end{minipage}
\begin{minipage}{0.245\textwidth}
  \rightline{\includegraphics[width=\textwidth]{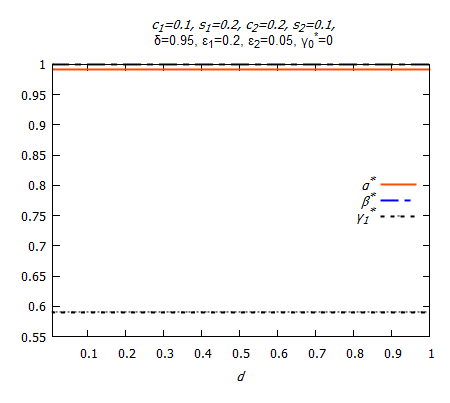}}
  \centerline{(\emph{e}) }
\end{minipage}
\begin{minipage}{0.245\textwidth}
  \leftline{\includegraphics[width=\textwidth]{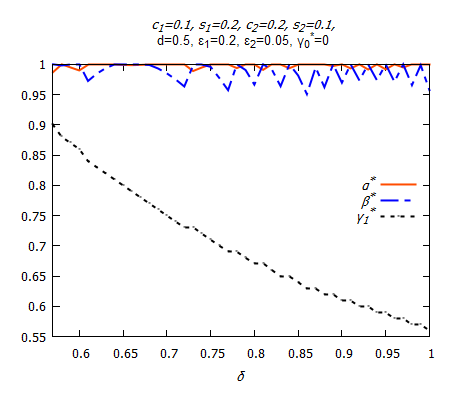}}
  \centerline{(\emph{f})}
\end{minipage}
\begin{minipage}{0.245\textwidth}
  \leftline{\includegraphics[width=\textwidth]{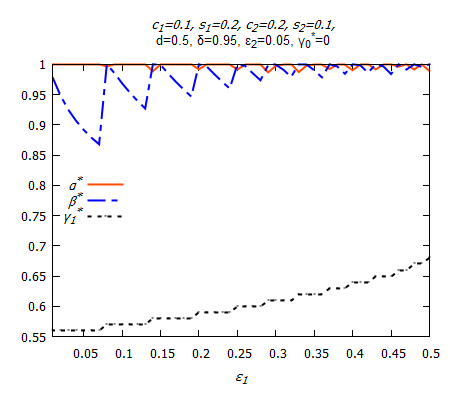}}
  \centerline{(\emph{g})}
\end{minipage}
\begin{minipage}{0.245\textwidth}
  \leftline{\includegraphics[width=\textwidth]{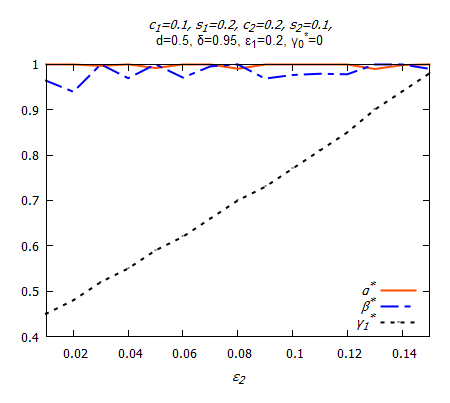}}
  \centerline{(\emph{h})}
\end{minipage}
 \caption{Design parameters versus intrinsic parameters: (a) $c_1$; (b) $c_2$; (c) $s_1$; (d) $s_2$; (e) $d$; (f) $\delta$; (g) $\varepsilon_1$; (h) $\varepsilon_2$.}
\label{fig:res}
\end{figure*}

In the first experiment, we illustrate how the optimal design $(\alpha^\ast,\beta^\ast,\gamma_1^\ast)$ is influenced by intrinsic parameters: (a) $c_1$, (b) $c_2$, (c) $s_1$, (d) $s_2$, (e) $d$, (f) $\delta$, (g) $\varepsilon_1$ and (h) $\varepsilon_2$ in Figure 3. The line chart plots the optimal value of $(\alpha^\ast,\beta^\ast,\gamma_1^\ast)$ in the optimal design rating protocol. In the remainder of this section, we set $s_1=0.2,c_2=0.2,s_2=0.1,d=0.5,\delta=0.95,\varepsilon_1=0.2$ and $\varepsilon_2=0.05$ by default without further notice.

Figure 3(a) illustrates the impact of $c_1$ on $(\alpha^\ast,\beta^\ast,\gamma_1^\ast)$. As shown by the results in Figure 3(a), $\gamma_1^\ast$ increases as $c_1$ increases when $c_1>c_2$. The major reason is that we only have to make sure worker 2 has enough incentive to comply with the recommend strategy as he has more desire to deviate $c_1<c_2$. Besides, a compliant worker's expected one-period utility monotonically decreases with $c_1$, and hence we have to increase the strength of incentive by increasing $\gamma_1$ to incentive workers to always behave well even with a large $c_1$. A similar phenomenon can be observed in Figure 3(b), $\gamma_1^\ast$ increases as $c_1$ increases when $c_2>c_1$.

Different from Figure 3(a) and 3(b), $\gamma_1^\ast$ remains unchanged as $s_1$ increases in Figure 3(c). This is due to the fact that $c_1<c_2$ and thus worker 2 prefers to deviate from the recommend strategy, if we ensure worker 2 abide by the recommend strategy, worker 1 certainly will not deviate even $s_1$ increases. However, in Figure 3(d), $\gamma_1^\ast$ increases as $s_2$ increases. This is because increasing the cost $s_2$ reduces the gap between $v_{\mathcal P}^\infty(CA)$ and $v_{\mathcal P}^\infty(CN)$, thereby leading to a reduced punishment on workers and we have to increase the strength of reward to incentive workers conform to the recommend strategy.

Similar to Figure 3(c), as $d$ increases, the $\gamma_1^\ast$ keeps the same as shown in Figure 3(e). The reason behind this phenomenon is the same. Figure 3(f) illustrates the impact of $\delta$ on design parameters. As $\delta$ increases, a worker puts a higher weight on his future utility relative to his instant utility, and hence, it is easy to give incentive to workers to comply with the social norm, thereby leading to a lower punishment (decreased $\gamma_1^\ast$) upon workers, and an increased optimal social welfare when workers have more patience. Figure 3(g) and 3(h) show that $\gamma_1^\ast$ increases as $\varepsilon_1$ (or $\varepsilon_2$) when $\varepsilon_2$ (or $\varepsilon_1$) is fixed, respectively. The reason is that the increase of $\varepsilon_1$ or $\varepsilon_2$ will cause a worker who comply with recommend strategy be mistakenly identified as a deviating worker, hence decreases his utility. Thereby we need a higher strength of incentive by increasing $\gamma_1^\ast$. Besides, the impact of $\varepsilon_2$ on $\gamma_1$ is greater than $\varepsilon_1$. This is due to the monitoring error happened in the second stage will caused more costs than monitoring error happened in the first stage. According to Figure 3 we can find that different $\gamma_1^\ast$ will cause different $\alpha^\ast$ and $\beta^\ast$ and this is corresponding to our theoretical result.

\subsection{Experiments on Optimal Social Utility Against Intrinsic Parameters}

In the second experiment, we examine the optimal social utility $U_{\mathcal P}$ against intrinsic parameters $c_1$, $c_2$, $s_1$, $s_2$, $d$, $\delta$, $\varepsilon_1$ and $\varepsilon_2$. The optimal value of social utility takes the maximum of these four cases (\emph{i.e.}, Cases $\romannumeral1$, $\romannumeral2$, $\romannumeral3$ and $\romannumeral4$).

As shown by the results in Figure 4(a), the social utility $U_{\mathcal P}$ decreases as $c_1$ increases when $c_1>c_2$. A similar phenomenon can be found in Figure 4(b) that $U_{\mathcal P}$ decreases as $c_2$ increases when $c_2>c_1$. This is because we need to increase the strength of incentive by increasing $\gamma_1^\ast$ as $c_1, c_2$ and $d$ increase to prevent the focal worker from deviating, if the cost of the focal worker in the fist stage greater than the opponent worker's. Different from Figure 4(a) and 4(b), $U_{\mathcal P}$ remains unchanged as $s_1$ increases as shown in Figure 4(c). The reason behind this phenomenon is the same as that of Figure 3(c), the requester does not need to increase reward to rase the strength of incentive when $c_1<c_2$. However, $U_{\mathcal P}$ increases as $s_2$ increases as shown in Figure 4(d), this is due to the fact that we need to increase the reward as $s_2$ increases, and hence the social utility decreases. In Figure 4(e), the social utility $U_{\mathcal P}$ remain unchanged as $d$ increases. This is because that workers will not deviate even $d$ increase since the exists of incentive mechanism. In Figure 4(f), the social utility $U_{\mathcal P}$ increases as $\delta$ increases. It is easy to give a worker with a higher $\delta$ incentive to follow the recommended strategies, thereby we can decrease $\gamma_1^\ast$ properly to receive higher social utility as $\delta$ increases. As shown in Figure 4(g) and 4(h), the social utility decreases as $\varepsilon_1$ and $\varepsilon_2$ increase. The main reason is the same as shown in Figure 3(g) and 3(h), since $\gamma_1^\ast$ increases as $\varepsilon_1$ and $\varepsilon_2$ increase, hence, the social utility will be decreased.

\begin{figure*}
\centering

\begin{minipage}{0.245\textwidth}
  \rightline{\includegraphics[width=\textwidth]{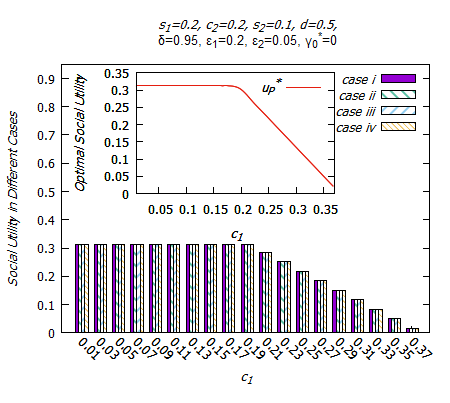}}
  \centerline{(\emph{a}) }
\end{minipage}
\begin{minipage}{0.245\textwidth}
  \leftline{\includegraphics[width=\textwidth]{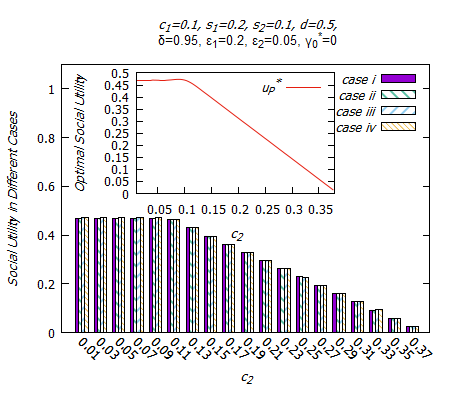}}
  \centerline{(\emph{b})}
\end{minipage}
\begin{minipage}{0.245\textwidth}
  \rightline{\includegraphics[width=\textwidth]{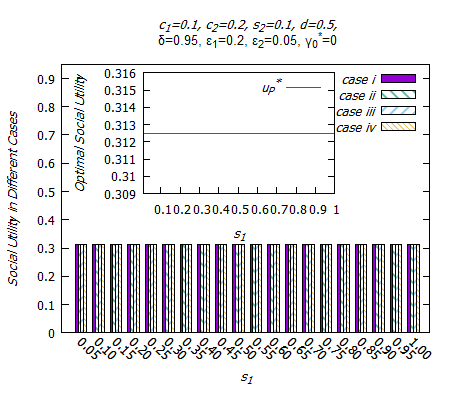}}
  \centerline{(\emph{c}) }
\end{minipage}
\begin{minipage}{0.245\textwidth}
  \leftline{\includegraphics[width=\textwidth]{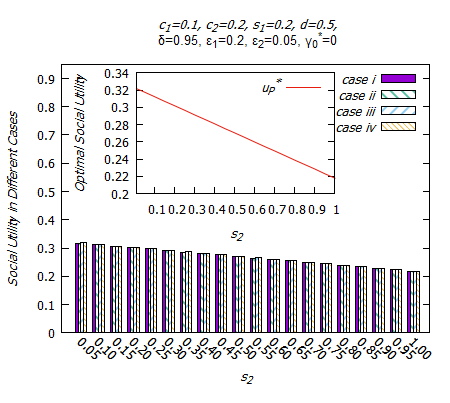}}
  \centerline{(\emph{d})}
\end{minipage}
\begin{minipage}{0.245\textwidth}
  \rightline{\includegraphics[width=\textwidth]{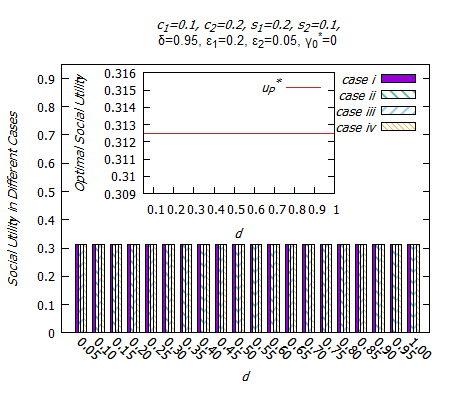}}
  \centerline{(\emph{e}) }
\end{minipage}
\begin{minipage}{0.245\textwidth}
  \leftline{\includegraphics[width=\textwidth]{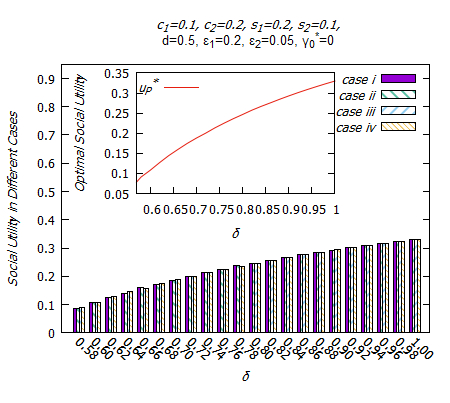}}
  \centerline{(\emph{f})}
\end{minipage}
\begin{minipage}{0.245\textwidth}
  \leftline{\includegraphics[width=\textwidth]{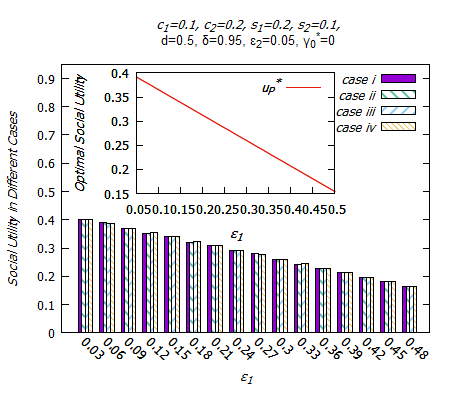}}
  \centerline{(\emph{g})}
\end{minipage}
\begin{minipage}{0.245\textwidth}
  \leftline{\includegraphics[width=\textwidth]{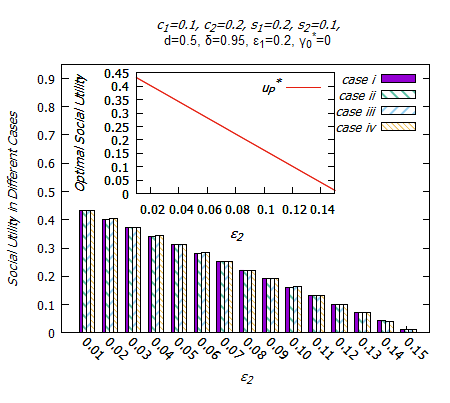}}
  \centerline{(\emph{h})}
\end{minipage}
 \caption{Social utility against intrinsic parameters: (a) $c_1$; (b) $c_2$; (c) $s_1$; (d) $s_2$; (e) $d$; (f) $\delta$; (g) $\varepsilon_1$; (h) $\varepsilon_2$. }
\label{fig:res4}
\end{figure*}

\section{CONCLUSION AND FUTURE WORK}

In this paper, we developed a novel game-theoretic design of rating protocols that integrated binary rating labels with differential pricing to address the crowdsourcing contest dilemma. In particular, we stood at the requester's point of view, and studied strategies for requesters to successfully maximize their utilities by extorting selfish workers and enforcing cooperation among them. By rigorously analyzing how workers' behaviors are influenced by intrinsic parameters, design parameters, as well as workers' evaluation of their individual long-term utilities, we characterize the optimal design by selecting four optimal design parameters $(\alpha^\ast,\beta^\ast,\gamma_1^\ast,\gamma_0^\ast)$, which are related to differential punishments and pricing schemes. Under the designed optimal rating protocol, requesters pay their workers as little as possible while enforcing sufficient incentives for individual workers to follow the social norm, even in the presence of imperfect monitoring.

In the future research, we intend to extend this work in several directions. Firstly, it is interesting to consider a large number of non-long-lived individuals in a crowdsourcing platform and aim to maximize the social welfare (\emph{i.e.}, the sum utility) of both workers and requesters. Secondly, we assume in this paper that each individual in the crowdsourcing platform has two roles, one as a requester and the other as a worker, and the role of an individual can not be switched in the next period. It is interesting to design appropriate rating protocols not following this assumption. Thirdly, we have assumed that individuals are entirely rational in crowdsourcing platform, however this assumption does not always hold in practice, and hence using real data to check the validity of proposed rating protocol is a next step for our future work.

\ifCLASSOPTIONcompsoc
  \section*{Acknowledgments}
\else
  \section*{Acknowledgment}
\fi

This work was supported in part by the National Natural Science Foundation of China under Grant 61402418, 11771013, 61751303, 11531011, 61503342, 61672468, 61602418, and in part by the Social development project of Zhejiang provincial public technology research under Grant 2017C33054.

 \section*{Appendix}
 \subsection{The Computation Process for Table \ref{table2}}
  \begin{table}[tb]
\caption{The pay-off matrix of each worker for the second-stage game under the (C,C) case}
\label{table3}
\begin{center}
\begin{tabular}{cc|c|c|c|c|c|c|}

  \multicolumn{2}{c}{\textbf{$$}}& \multicolumn{6}{c}{$worker$ 2} \\
  \multicolumn{2}{c}{\textbf{$$}}& \multicolumn{3}{c}{\textbf{$A$}}& \multicolumn{3}{c}{\textbf{$N$}} \\
\cline{3-8}
\multirow{2}*{$worker$ 1} & \multicolumn{1}{c}{\textbf{$A$}}& \multicolumn{3}{|c|}{$1-c_1-s_1,-c_2-s_2$}& \multicolumn{3}{c|}{$1-c_1-s_1,-c_2$} \\
\cline{3-8}
\multicolumn{1}{c}{\textbf{$$}}& \multicolumn{1}{c}{\textbf{$N$}}& \multicolumn{3}{|c|}{$-c_1,1-c_2-s_2$}& \multicolumn{3}{c|}{$1-c_1,-c_2$} \\
\cline{3-8}
\end{tabular}%
\end{center}
\end{table}

 \begin{table}[tb]
\caption{The pay-off matrix of each worker for the second-stage game under the (S,S) case}
\label{table4}
\begin{center}
\begin{tabular}{cc|c|c|c|c|c|c|}

  \multicolumn{2}{c}{\textbf{$$}}& \multicolumn{6}{c}{$worker$ 2} \\
  \multicolumn{2}{c}{\textbf{$$}}& \multicolumn{3}{c}{\textbf{$A$}}& \multicolumn{3}{c}{\textbf{$N$}} \\
\cline{3-8}
\multirow{2}*{$worker$ 1} & \multicolumn{1}{c}{\textbf{$A$}}& \multicolumn{3}{|c|}{$1-s_1,-s_2$}& \multicolumn{3}{c|}{$1-s_1,0$} \\
\cline{3-8}
\multicolumn{1}{c}{\textbf{$$}}& \multicolumn{1}{c}{\textbf{$N$}}& \multicolumn{3}{|c|}{$0,1-s_2$}& \multicolumn{3}{c|}{$1,0$} \\
\cline{3-8}
\end{tabular}%
\end{center}
\end{table}

 \begin{table}[tb]
\caption{The pay-off matrix of each worker for the second-stage game under the (C,S) case}
\label{table5}
\begin{center}
\begin{tabular}{cc|c|c|c|c|c|c|}

  \multicolumn{2}{c}{\textbf{$$}}& \multicolumn{6}{c}{$worker$ 2} \\
  \multicolumn{2}{c}{\textbf{$$}}& \multicolumn{3}{c}{\textbf{$A$}}& \multicolumn{3}{c}{\textbf{$N$}} \\
\cline{3-8}
\multirow{2}*{$worker$ 1} & \multicolumn{1}{c}{\textbf{$A$}}& \multicolumn{3}{|c|}{$1-c_1-s_1,-s_2$}& \multicolumn{3}{c|}{$1-c_1-s_1,0$} \\
\cline{3-8}
\multicolumn{1}{c}{\textbf{$$}}& \multicolumn{1}{c}{\textbf{$N$}}& \multicolumn{3}{|c|}{$1-c_1-d,-(s_2+d)(1-s_2)$}& \multicolumn{3}{c|}{$1-c_1,0$} \\
\cline{3-8}
\end{tabular}%
\end{center}
\end{table}

 \begin{table}[tb]
\caption{The pay-off matrix of each worker for the second-stage game under the (S,C) case}
\label{table6}
\begin{center}
\begin{tabular}{cc|c|c|c|c|c|c|}

  \multicolumn{2}{c}{\textbf{$$}}& \multicolumn{6}{c}{$worker$ 2} \\
  \multicolumn{2}{c}{\textbf{$$}}& \multicolumn{3}{c}{\textbf{$A$}}& \multicolumn{3}{c}{\textbf{$N$}} \\
\cline{3-8}
\multirow{2}*{$worker$ 1} & \multicolumn{1}{c}{\textbf{$A$}}& \multicolumn{3}{|c|}{$-s_1,1-c_2-s_2$}& \multicolumn{3}{c|}{$0,1-c_2-s_2$} \\
\cline{3-8}
\multicolumn{1}{c}{\textbf{$$}}& \multicolumn{1}{c}{\textbf{$N$}}& \multicolumn{3}{|c|}{$-(s_1+d)(1-s_1),1-c_2-d$}& \multicolumn{3}{c|}{$0,1-c_2$} \\
\cline{3-8}
\end{tabular}%
\end{center}
\end{table}

The concrete computation process for Table \ref{table2} in four cases is shown as follows:

Case $\uppercase\expandafter{\romannumeral 1}$: (\emph{C,C}), \emph{i.e.,} both workers choose \emph{C}. The pay-off matrix showing utilities for Case (\emph{C,C}) appears in Table \ref{table3}. Let $p_1$ and $p_2$ denote productivities of worker 1 and worker 2, respectively, which are known before they decide on attacking. The ex-ante utility of worker 1 is
\begin{equation} {}
   \begin{split}
   v_1=&Pr(p_2<p_1<p_2+d)(1-c_1-s_1)+\\
   &Pr(p_1>p_2+d)(1-c_1)\\
   =&1/2-c_1/2-s_1d+(s_1d^2)/2
   \end{split}
   \end{equation}
And the ex-ante utility of worker 2 is
\begin{equation} {}
   \begin{split}
   v_2=&Pr(p_1<p_2<p_1+d)(1-c_2-s_2)+\\
   &Pr(p_2>p_1+d)(1-c_2)\\
   =&1/2-c_2/2-s_2d+(s_2d^2)/2
   \end{split}
   \end{equation}

Case $\uppercase\expandafter{\romannumeral 2}$: (\emph{S,S}), \emph{i.e.,} both worker 1 and worker 2 choose S. The pay-off matrix showing utilities for Case (\emph{S,S}) appears in Table \ref{table4}. The ex-ante utility of worker 1 is
\begin{equation} {}
   \begin{split}
   v_1&=Pr(p_2<p_1<p_2+d)(1-s_1)+Pr(p_1>p_2+d)\\
   &=1/2-s_1d+(s_1d^2)/2
   \end{split}
   \end{equation}
And the ex-ante utility of worker 2 is
\begin{equation} {}
   \begin{split}
   v_2&=Pr(p_1<p_2<p_1+d)(1-s_2)+Pr(p_2>p_1+d)\\
   &=1/2-s_2d+(s_2d^2)/2
   \end{split}
   \end{equation}

Case $\uppercase\expandafter{\romannumeral 3}$: (\emph{C,S}), \emph{i.e.,} worker 1 chooses \emph{C} and worker 2 chooses \emph{S}. The pay-off matrix showing utilities for the case of (\emph{C,S}) appears in Table \ref{table5}. The ex-ante utility of worker 1 is
\begin{equation} {}
   \begin{split}
   v_1&=Pr(p_2+c_1<p_1<p_2+c_1+d)(1-c_1-s_1)\\
   &~~~~+Pr(p_1>p_2+c_1+d)(1-c_1)\\
   &=[d-(c_1+d)^2/2](1-c_1-s_1)\\
   &~~~~+(1-c_1-d)^2(1-c_1)/2
   \end{split}
   \end{equation}

And the ex-ante utility of worker 2 is
\begin{equation} {}
   \begin{split}
   &v_2=0
   \end{split}
   \end{equation}

Case $\uppercase\expandafter{\romannumeral 4}$: (\emph{S,C}), \emph{i.e.,} worker 1 chooses \emph{S} and worker 2 chooses \emph{C}. The pay-off matrix showing utilities for Case (\emph{C,S}) appears in Table \ref{table6}. The ex-ante utility of worker 1 is
\begin{equation} {}
   \begin{split}
   &v_1=0
   \end{split}
   \end{equation}
And the ex-ante utility of worker 2 is
\begin{equation} {}
   \begin{split}
   v_2&=Pr(p_1+c_2<p_2<p_1+c_2+d)(1-c_2-s_2)\\
   &~~~~+Pr(p_2>p_1+c_2+d)(1-c_2)\\
   &=[d-(c_2+d)^2/2](1-c_2-s_2)\\
   &~~~~+(1-c_2-d)^2(1-c_2)/2
   \end{split}
   \end{equation}

 \subsection{Utilities}
 \begin{table*}[tb]

\caption{THE UTILITY MATRIX OF ONE TRANSACTION UNDER PERFECT MONITORING}
\label{table7}
\begin{center}
\begin{tabular}{cc|c|c|c|c|c|ccc}

  \multicolumn{2}{c}{\textbf{$$}}& \multicolumn{8}{c}{$worker$ 2} \\
  \multicolumn{2}{c}{\textbf{$$}}& \multicolumn{2}{c}{\textbf{$CN$}}& \multicolumn{2}{c}{\textbf{$CA$}} & \multicolumn{2}{c}{\textbf{$SN$}} & \multicolumn{2}{c}{\textbf{$SA$}} \\
\cline{3-10}
\multirow{8}*{$worker$ 1}
& \multirow{2}*{$CN$}&
\multicolumn{2}{c|}{$\gamma/2-c_1$}& \multicolumn{2}{c|}{$-c_1-d$}& \multicolumn{2}{|c|}{$\gamma-c_1$}& \multicolumn{2}{c|}{$\gamma-c_1-d$} \\
\multicolumn{1}{c}{\textbf{$$}}& \multicolumn{1}{c}{\textbf{$$}}& \multicolumn{2}{|c|}{$\gamma/2-c_2$}& \multicolumn{2}{c|}{$\gamma-c_2-s_2$}& \multicolumn{2}{c|}{$0$} & \multicolumn{2}{c|}{$-s_2$} \\
\cline{3-10}

& \multirow{2}*{$CA$}&
\multicolumn{2}{c|}{$\gamma-c_1-s_1$}& \multicolumn{2}{c|}{$\gamma/2-c_1-s_1-d$}& \multicolumn{2}{|c|}{$\gamma-c_1-s_1$}& \multicolumn{2}{c|}{$\gamma-c_1-s_1-d$} \\
\multicolumn{1}{c}{\textbf{$$}}& \multicolumn{1}{c}{\textbf{$$}}& \multicolumn{2}{|c|}{$-c_2-d$}& \multicolumn{2}{c|}{$\gamma/2-c_2-s_2-d$}& \multicolumn{2}{c|}{$-d$} & \multicolumn{2}{c|}{$-s_2-d$} \\
\cline{3-10}

& \multirow{2}*{$SN$}&
\multicolumn{2}{c|}{$0$}& \multicolumn{2}{c|}{$-d$}& \multicolumn{2}{|c|}{$\gamma/2$}& \multicolumn{2}{c|}{$-d$} \\
\multicolumn{1}{c}{\textbf{$$}}& \multicolumn{1}{c}{\textbf{$$}}& \multicolumn{2}{|c|}{$\gamma-c_2$}& \multicolumn{2}{c|}{$\gamma-c_2-s_2$}& \multicolumn{2}{c|}{$\gamma/2$} & \multicolumn{2}{c|}{$\gamma-s_2$} \\
\cline{3-10}

& \multirow{2}*{$SA$}&
\multicolumn{2}{c|}{$-s_1$}& \multicolumn{2}{c|}{$-s_1-d$}& \multicolumn{2}{|c|}{$\gamma-s_1$}& \multicolumn{2}{c|}{$\gamma/2-s_1-d$} \\
\multicolumn{1}{c}{\textbf{$$}}& \multicolumn{1}{c}{\textbf{$$}}& \multicolumn{2}{|c|}{$\gamma-c_2-d$}& \multicolumn{2}{c|}{$\gamma-c_2-s_2-d$}& \multicolumn{2}{c|}{$-d$} & \multicolumn{2}{c|}{$\gamma/2-s_2-d$} \\
\cline{3-10}

\end{tabular}%
\end{center}
\end{table*}

\begin{figure*}
\begin{equation}
V_i=\left[
\begin{matrix}
\gamma/2-c_i&-c_i-d&\gamma-c_i&\gamma-c_i-d\\
\gamma-c_i-s_i&\gamma/2-c_i-s_i-d&\gamma-c_i-s_i&\gamma-c_i-s_i-d\\
0&-d&\gamma/2&-d\\
-s_i&-s_i-d&\gamma-s_i&\gamma/2-s_i-d
\end{matrix}
\right]
\end{equation}
\end{figure*}

We first give the pay-off matrix which shows the utility of each worker in one-period under perfect monitoring, where the total reward is $\gamma$. The utility matrix of one-period under perfect monitoring is in TABLE \ref{table7}. The expected payoff of $i$th-worker is determined by his adopted actions, his current rating label as well as his opponent under the monitoring error probabilities $\varepsilon_1$ and $\varepsilon_2$, which can be derived as follows:
\begin{eqnarray}{}
\label{payoff1}
\left \{
\begin{array}{ll}
  v_{CN}^i(\theta,\tilde{\theta})=A\times V_i\times A^T \\
  v_{CA}^i(\theta,\tilde{\theta})=B\times V_i\times A^T \\
  v_{SN}^i(\theta,\tilde{\theta})=C\times V_i\times A^T \\
  v_{SA}^i(\theta,\tilde{\theta})=D\times V_i\times A^T
\end{array}
\right.
\end{eqnarray}

The array $V_i$ is directly derived from Table \ref{table7}, which is shown at the top of the next page. The order of strategies for the assigned probabilities in \emph{A}, \emph{B}, \emph{C} and \emph{D} are \emph{CN}, \emph{CA}, \emph{SN} and \emph{SA} as shown in the following, respectively.
\begin{eqnarray}{}
      \left \{
\begin{array}{ll}
  A=[(1-\varepsilon_1)(1-\varepsilon_2),(1-\varepsilon_1)\varepsilon_2,\varepsilon_1(1-\varepsilon_2),\varepsilon_1\varepsilon_2] \\
  B=[(1-\varepsilon_1)\varepsilon_2,(1-\varepsilon_1)(1-\varepsilon_2),\varepsilon_1\varepsilon_2,\varepsilon_1(1-\varepsilon_2)] \\
  C=[\varepsilon_1(1-\varepsilon_2),\varepsilon_1\varepsilon_2,(1-\varepsilon_1)(1-\varepsilon_2),(1-\varepsilon_1)\varepsilon_2] \\
  D=[\varepsilon_1\varepsilon_2,\varepsilon_1(1-\varepsilon_2),(1-\varepsilon_1)\varepsilon_2,(1-\varepsilon_1)(1-\varepsilon_2)]
\end{array}
\right.
\end{eqnarray}

Given a fixed pricing scheme $\psi$ under the proposed rating protocol $\mathcal P$, there exists a unique stationary rating distribution $\{\eta_{\mathcal P}^t(\theta)\}$ as shown in Eq.(\ref{distribution}), then Eq.(\ref{payoff1}) can be rewritten as
\begin{eqnarray}{}
\label{payoff2}
\left \{
\begin{array}{ll}
  v_{CN}^i(\theta)=\sum_{\tilde{\theta}=0}^1 \eta(\tilde{\theta})v_{CN}^i(\theta,\tilde{\theta}) \\
  v_{CA}^i(\theta)=\sum_{\tilde{\theta}=0}^1 \eta(\tilde{\theta})v_{CA}^i(\theta,\tilde{\theta}) \\
  v_{SN}^i(\theta)=\sum_{\tilde{\theta}=0}^1 \eta(\tilde{\theta})v_{SN}^i(\theta,\tilde{\theta}) \\
  v_{SA}^i(\theta)=\sum_{\tilde{\theta}=0}^1 \eta(\tilde{\theta})v_{SA}^i(\theta,\tilde{\theta})
\end{array}
\right.
\end{eqnarray}

\ifCLASSOPTIONcaptionsoff
  \newpage
\fi

\end{document}